\documentclass[%
reprint,
amsmath,amssymb,
aps,
]{revtex4-2}
\usepackage[toc,page]{appendix}
\usepackage[dvipsnames]{xcolor}
\usepackage{gensymb}
\usepackage{lipsum, babel}
\usepackage{graphicx}
\usepackage{dcolumn}
\usepackage{bm}
\usepackage{amsmath}
\usepackage{physics}
\usepackage[utf8]{inputenc}
\usepackage{tablefootnote}
\usepackage{soul}
\usepackage[normalem]{ulem}

\usepackage{chemformula} 
\usepackage[colorlinks=true,allcolors=blue]{hyperref}

\usepackage{graphicx}
\usepackage{xcolor} 
\begin{document}
	
\preprint{APS/123-QED}
	
\title{Leveraging Plasmonic Hot Electrons to Quench Defect Emission in Metal - Semiconductor Nanostructured Hybrids: Experiment and Modeling}

\author{Kritika Sharu}
\email{kritika.iiser16@iisertvm.ac.in}
\affiliation{School of Physics, Indian Institute of Science Education and Research Thiruvananthapuram, Kerala 695551, India}

\author{Shashwata Chattopadhyay }
\affiliation{School of Physics, Indian Institute of Science Education and Research Thiruvananthapuram, Kerala 695551, India}

\author{K. N. Prajapati}
\affiliation{School of Physics, Indian Institute of Science Education and Research Thiruvananthapuram, Kerala 695551, India}

\author{J. Mitra }
\email{j.mitra@iisertvm.ac.in}
\affiliation{School of Physics, Indian Institute of Science Education and Research Thiruvananthapuram, Kerala 695551, India}

\begin{abstract}
Modeling light-matter interaction in hybrid plasmonic materials is vital to their widening relevance from optoelectronics to photocatalysis. Here, we explore photoluminescence from ZnO nanorods (ZNR) embedded with gold nanoparticles (Au NPs). 
Progressive increase in Au NP concentration introduces significant structural disorder and defects in the ZNRs, which paradoxically quenches defect related visible photoluminescence (PL) while intensifying the near band edge (NBE) emission. Under UV excitation, the simulated semi-classical model realizes PL from ZnO with sub-band gap defect states, eliciting visible emissions that are absorbed by Au NPs to generate a non-equilibrium hot carrier distribution. The photo-stimulated hot carriers, transferred to ZnO, substantially modify its steady-state luminescence, reducing NBE emission lifetime and altering the abundance of ionized defect states, finally reducing visible emission. The simulations show that the change in the interfacial band bending at the Au-ZnO interface under optical illumination facilitates charge transfer between the components. This work provides a general foundation to observe and model the hot carrier dynamics in hybrid plasmonic systems.   
		
\end{abstract}

\maketitle

\section{Introduction}
Surface plasmon resonances (SPR) of metal nanostructures finds wide-ranging applications stemming from the capability to concentrate and enhance light energy into ultra-small sub-wavelength volumes and their large scattering cross-section in the visible \cite{kalathingal2016scanning,RN719, RN612, Wu2015, RN717, RN656,PhysRevLett.108.213907}. 
The localized surface plasmons (LSP) and the polaritons de-excite radiatively \cite{RN575,Mertens2017a} or, more often non-radiatively through electron-electron, electron-phonon and defect or surface scattering \cite{Brown2016,Pinchuk_2003,RN694}, which has been exploited for localized heating\cite{RN695}. Plasmons may also decay via secondary electron-hole ($\textit{e-h}$) pair generation, forming an uncorrelated, non-equilibrium hot carrier distribution. The distribution is commensurate with the plasmon energy ($\hbar\omega_{pl}$) and characterized by a temperature ($T_e$) that is significantly higher than the equilibrium lattice temperature ($T_{eq}$) \cite{RN696}. There is significant interest in exploiting hot carrier dynamics in an emerging class of hybrid plasmonic systems composed of plasmonic nanostructures and semiconductors. The emission properties of semiconducting emitters often benefit from overlap in its spectral response with plasmonic field enhancement\cite{C4CS00131A}. However, even in the off-resonant regime, hybrid systems can be designed to leverage hot carriers to tailor the spectral characteristics and improve efficiency of photovoltaic devices and photocatalysts\cite{clavero2014plasmon,li2017harvesting,Linic2021,mali2016situ}.    

 Here, we explore a hybrid plasmonic system of ZnO nanorods (ZNR) embedded with Au nanoparticles (NP). ZnO, a wide band gap ($E_g \sim$ 3.3 eV) semiconductor, has been widely explored in conjunction with plasmonic NPs to achieve luminescence tuning, enhanced sensor performance\cite{Chen_2008,KIM2021658, RN692}, and improved photocatalytic activity \cite{RN691,CASTELLOLUX2023118377,chiu2018plasmon}. The typical photoluminescence (PL) spectrum of ZnO shows a distinct near band edge (NBE)  excitonic peak around 370 nm, along with broad emission in the visible, that is attributed to interstitial Zn ($I_{Zn}$) and deep level (DL) oxygen vacancy ($V_O$) related defects\cite{Bandopadhyay2015d}. Consequently, the intensity and spectral composition of visible emission is highly dependent on the growth conditions and  morphology of the nanostructures and are susceptible to environmental conditions \cite{KANG2021115544}. The PL is also influenced by the highly inhomogeneous distribution of the neutral and ionized defect states ($V_O, V_O^+, V_O^{++}$), especially at interfaces and grain boundaries \cite{bandopadhyay2016spatially}. 
 Previous investigations from our group \cite{Prajapati2020d} and others \cite{RN720, RN713, RN714, RN715, RN716, khan2019enhanced, wang2021enhanced, guidelli2015enhanced} have shown that ZnO nanostructures coupled with plasmonic NPs lead to decay in visible PL, simultaneously enhancing NBE emission, a similar observation has been reported in Ag-GaN hybrids \cite{singh2023photocurrent} which have been exploited for novel applications in photovoltaics and spectroscopy\cite{Prajapati_2022}. 
 While the reports recognize strong coupling between the components that enable charge \& energy transfer, the investigations do not elucidate the role of the Au-ZnO Schottky interface, especially under photoexcitation that influences carrier exchange\cite{D2NR05937A}. Reduction of barrier width comparable to hot electron mean free path in the NPs \cite{RN693, RN708} and embedding NPs within the semiconductor  have been shown to increase charge transport efficiency\cite{RN712}, highlighting the importance of not only the choice of materials but also the  geometry of inclusion.  Effective engineering of hybrid systems and exploitation of their novelties also requires comprehension of the multiple concurrent physical  processes that enrich the response of such hybrids. Previous theoretical investigations have modeled PL in wide band gap systems and their variation with experimental control parameters, both in the steady state\cite{RN697,RN698,RN699,Li_2017} as well as their temporal response\cite{RN706}.  Further, the proximity effect of plasmonic NPs on semiconductor emitters, modifying their decay rates have been variously modeled to elucidate the role of emitter-plasmon coupling \cite{RN702}.   
However, these investigations omit the concurrence of several energetic and dynamical processes that plasmonic nanostructures induce in hybrid systems,  finally modifying the emission properties of the semiconductor. Our investigation is also motivated by a paucity of models with well-delineated parameters that simulate optical processes in the constituent materials and further offer physically motivated coupling that determines and controls the PL of the hybrid system. 

\begin{figure}
		\centering
		\includegraphics[width=0.9\linewidth]{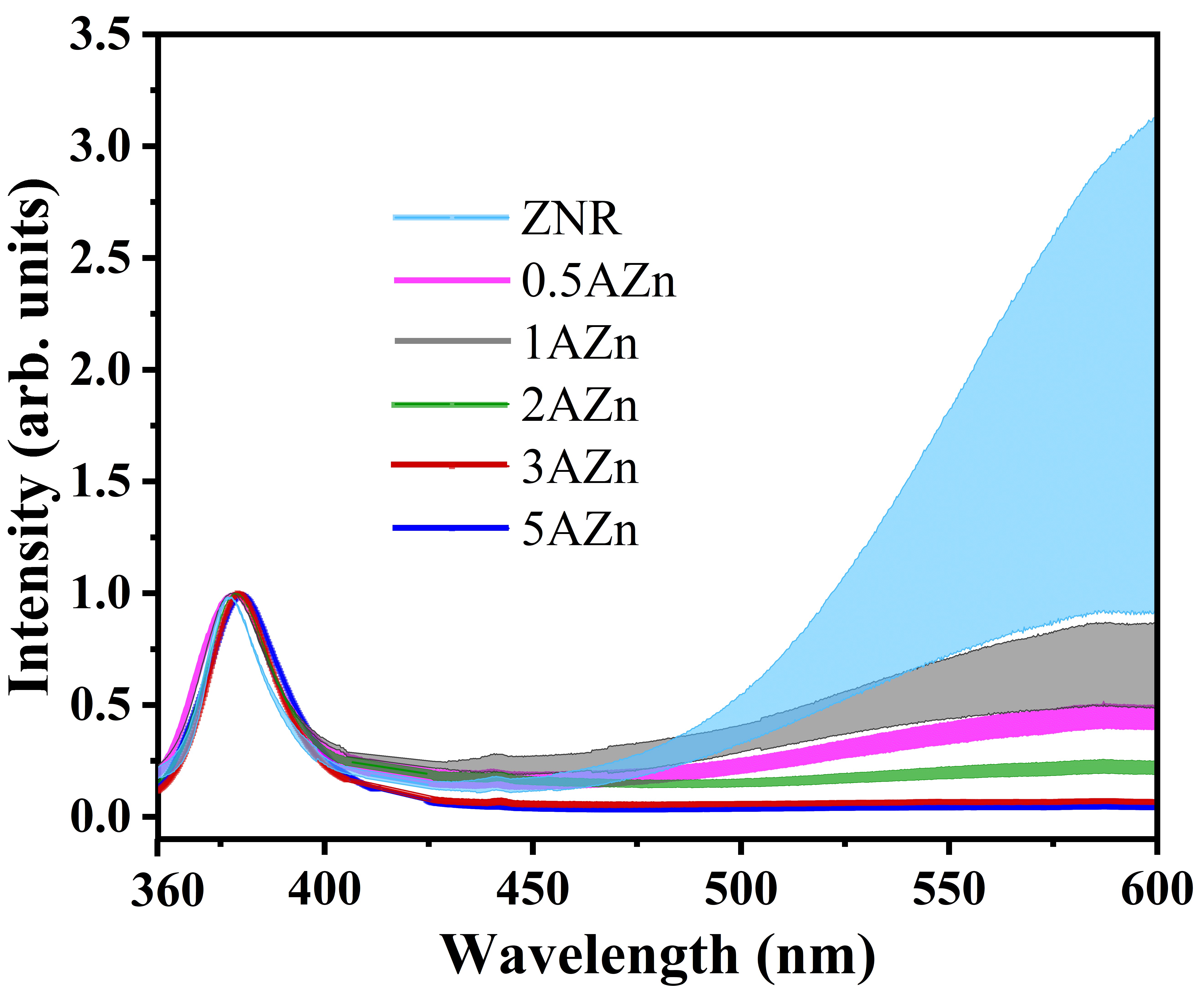}
		\caption{PL spectra of bare ZNR and Au NP embedded ZNRs with increasing concentration of Au NPs. Spectra are normalized with near band edge emission intensity and the spread denotes data across multiple samples.}
		\label{fig:PL_ALL}
\end{figure}

 Here, Au NPs embedded in the ZnO matrix create the most intimate and pristine contact between the two materials and modify the PL spectrum as shown in  Fig.\ref{fig:PL_ALL}. PL spectra of bare ZNR samples prepared under {\it identical} conditions show wide variation, especially in visible emission. 
 Spectra from samples with increasing  Au NP inclusion lead to progressive quenching of visible emission, accompanied by NBE emission enhancement.
The results are counter intuitive since the ZNRs become morphologically disordered with increasing NP incorporation, nucleating increasing density of native defects that would aid visible emission. 
We present a combined experimental and theoretical study to explore the spectral response of this hybrid plasmonic system with increasing Au NP inclusion in ZnO. 
Finite element method (FEM) modeling is used to implement a semiclassical  calculation to obtain recombination rates for bare ZNRs that are later augmented by modeling energy and charge transfer processes arising due to Au NP inclusion. Under UV illumination, the model evinces change to the metal-semiconductor interfacial band alignment which, along with an estimate for hot carrier density generated in the NPs, quantifies electron transfer from Au to ZnO. The results show direct increase in the interband transition rate and also reveal increased electron occupancy  of the DL ionized defect states in ZnO, due to electron transfer. This indirect occupancy modulation leads to the observed quenching in visible emission. Diverse applicability of hybrid plasmonic systems contextualizes the relevance of the present exercise in modeling their optical properties that may be extended to provide insights into the properties and parameters which govern their response.
	
\section{Experimental and Model details}
	All chemicals, including zinc nitrate hexahydrate, hexamethylene-tetramine (HMTA), Au NPs of mean diameter $\sim$ 30 ($\pm 20$) nm, were supplied by Sigma Aldrich. 
	All  ZNR samples were grown by a low-temperature hydrothermal route\cite{Bandopadhyay2015d} on Zn foil substrate, as detailed in Supplemental Material (SM) Section S1. A process with slow ZNR growth kinetics was chosen to promote the incorporation and retention of Au NPs within the ZnO lattice. To embed AuNP within the ZNR, a suspension of AuNPs in a stabilizing buffer ($\sim 10^{11}$ particles/ml) was added to the growth solution. SM table S1 lists the Au:Zn atomic ratio in the growth solution. By simultaneously increasing the AuNP solution volume and decreasing the zinc nitrate hexahydrate+HMTA stock solution volume, we obtain samples ranging from least AuNP inclusion(bare ZNR) to maximum inclusion(5AZn).  

Morphological characterization was performed using a field-emission scanning electron microscope (SEM, Nova Nano SEM 450) and transmission electron microscope (TEM, TECHNAI G2-TF-30). Structural analysis was carried out using a powder x-ray diffractometer (Empyrean, PANalytical) with reference radiation of Cu K$_\alpha$ = 1.540 Å. X-ray photo-electron spectroscopy (XPS) was done using a Scienta Omicron XPS system. The Raman spectra were recorded  with 638 nm and 532 nm excitation (Horiba Scientific Xplora Plus). PL spectra were recorded using a custom made setup using a spectrometer (Shamrock 303i and iDus 420 CCD, Oxford Instruments) under 320 nm laser excitation. All numerical calculations were performed using codes developed in Mathematica, python and finite element method (FEM) calculation software COMSOL Multiphysics 5.3a. Details of the model parameters are available in the SM Sections S3, S4 and S5. The optical properties of Au used are given by Rakic et al. \cite{rakic1998optical}. ZnO is modeled as an $n$-type direct band gap semiconducting dielectric with $E_g$ = 3.4 eV, electron affinity $\chi_e$ = 4.29 eV,  dopant density $\sim 10^{12}$/cc and relative permittivity $\epsilon_r$ = 8.3.

\section{Results and Discussion}
\begin{figure}
		\includegraphics[width=0.9\linewidth]{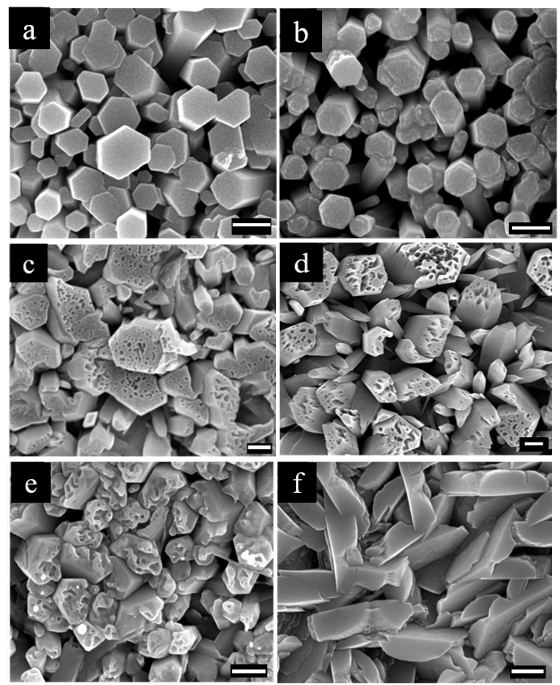}
		\caption{SEM Images: (a) ZNR, (b) 0.5AZn, (c) 1AZn, (d) 2AZn, (e) 3AZn, (f) 5AZn. All scale bars = 300 nm.} \label{Fig:SEM}
\end{figure}
SEM images of bare ZNR and Au-ZNR with increasing concentrations of Au NPs are shown in Fig. \ref{Fig:SEM} (a)-(f). The bare ZNRs show typical hexagonal nanorods with an average diameter of $300\ (\pm 100$) nm and uniform coverage across the Zn substrate. In contrast, Au-ZNR samples display distortion of the hexagonal shape and are riddled with pores (fig 1b-1e), with pore size and density increasing with increasing Au NP concentration. The pore size varies from 20 – 80 nm, which is comparable to the Au NP diameter. At the highest NP concentration (5AZn), the hexagonal shape is completely lost (Fig. \ref{Fig:SEM}(f)). 
\begin{figure}
	\centering
	\includegraphics[width=0.8\linewidth]{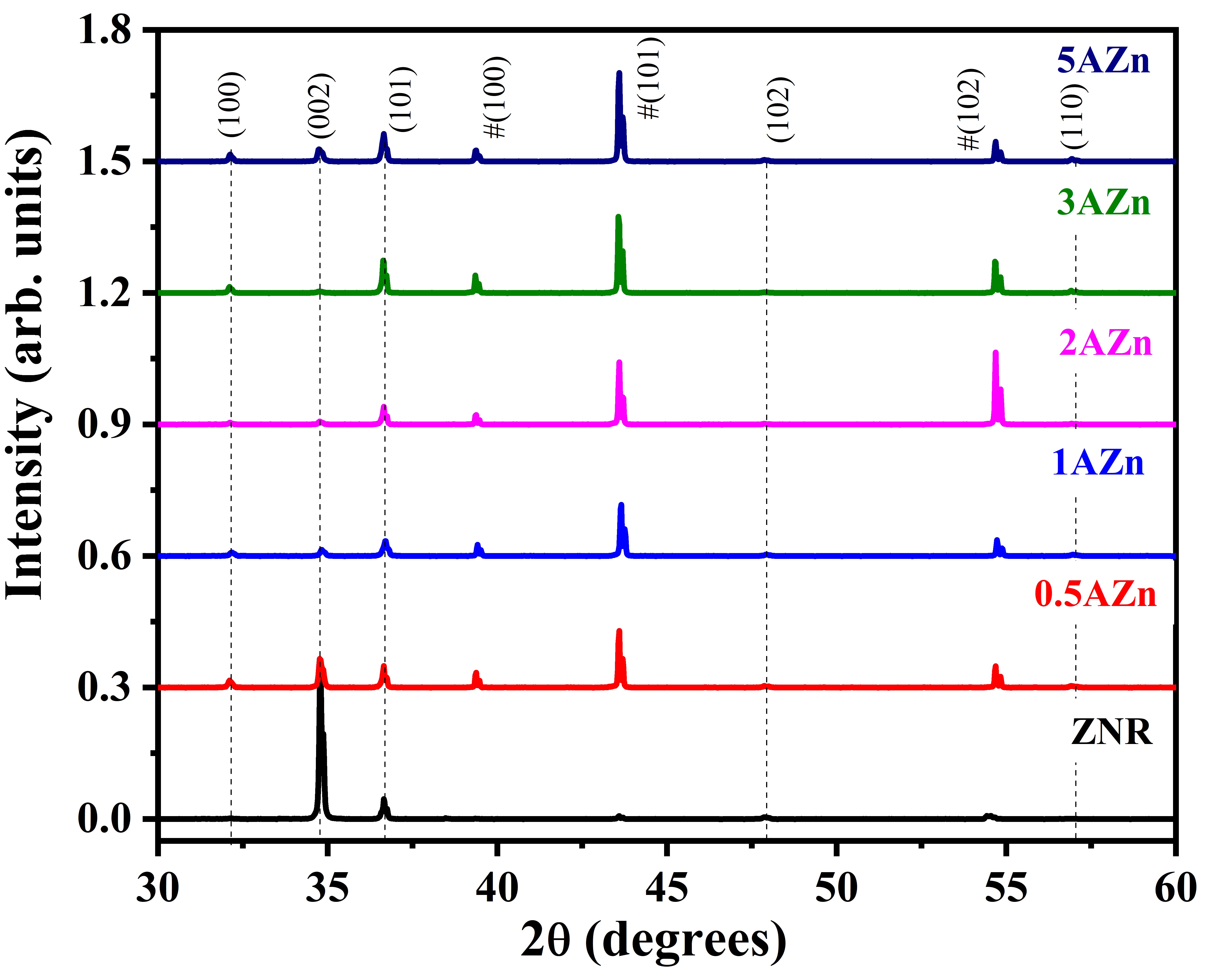}
	\caption{XRD pattern of AuNP concentration dependent XRD spectra, beginning from ZNR(bottom) to 5-AZn(topmost).} \label{Fig:XRD}
\end{figure}
The evolution in morphology is reflected in the XRD spectra of the samples, shown in Fig. \ref{Fig:XRD}. The spectrum for the bare ZNRs shows a dominant ZnO (002) wurtzite peak along with less intense (100), (101) peaks and signatures of the Zn foil are marked with $\#$.
\begin{figure}[b]
	\centering
	\includegraphics[width=\linewidth]{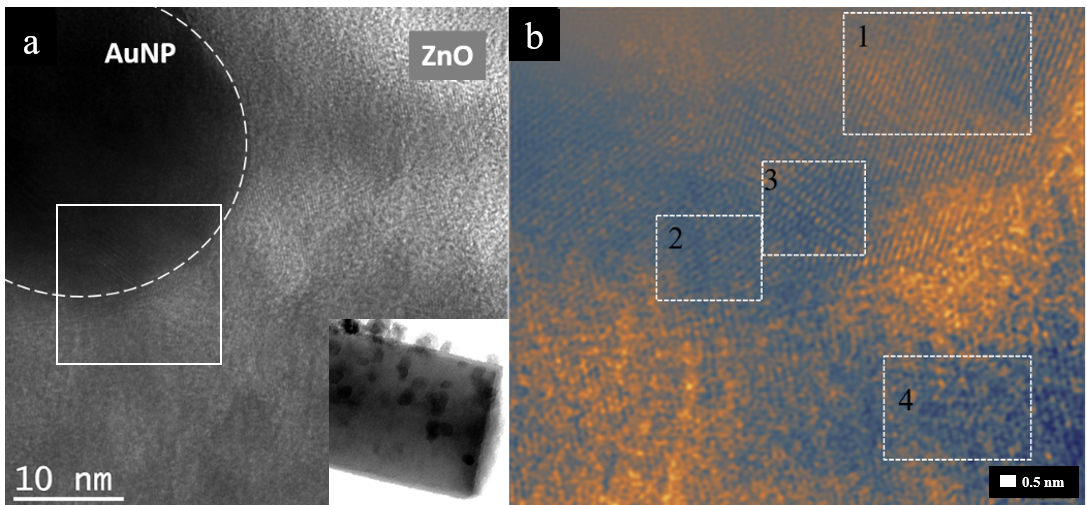}
	\caption{TEM images: (a) High resolution (HRTEM) image of an Au nanoparticle embedded in ZnO (Inset: Low resolution TEM image of AZn sample. (b) Color contrast enhanced image of region within the white box in (a).}
	\label{Fig:TEM}
\end{figure} The spectra for AZn samples show increasingly impeded c-axis growth of the ZNRs and emerging signature of (100), (101), (102) and (110) planes of the ZnO lattice. It appears that the incorporation of each Au NP hinders the local c-axis growth of the ZnO crystallites, leading to the creation of pores, compromising the pristine hexagonal morphology. Low-resolution TEM images of an AZn sample, shown in Fig. \ref{Fig:TEM}(a), image the presence of Au NPs (dark regions of diameter ~20-40 nm) on the surface and embedded within the ZNRs. Fig.\ref{Fig:TEM} (b) shows the HRTEM image of an AZn sample. The atomic planes of the nanorods and Au NPs are clearly evidenced by the fringe patterns. The c-axis oriented ZNRs shows commensurate planar spacing in the Fig.\ref{Fig:TEM} (b), simultaneously resolving the Au (111) planes of the AuNP, exhibiting fringe spacing of 0.28 and 0.22 nm, respectively. Fig.\ref{Fig:TEM} (c) shows magnified, contrast-enhanced  image of the Au - ZnO interface evidencing the crossover from the Au to ZnO lattice. Regions 1 and 2 of Au NP show fringe width 0.23 nm while region 3 shows spacing between 0.22 nm to 0.24 nm. Region 4 shows fringe width 0.28 nm corresponding to the (002) plane of ZNR. The absence of any signature of Au lattice in the XRD data is perhaps due to the very low concentration of Au NPs in the system.
	\begin{figure}
		\centering
		\includegraphics[width=0.9\linewidth]{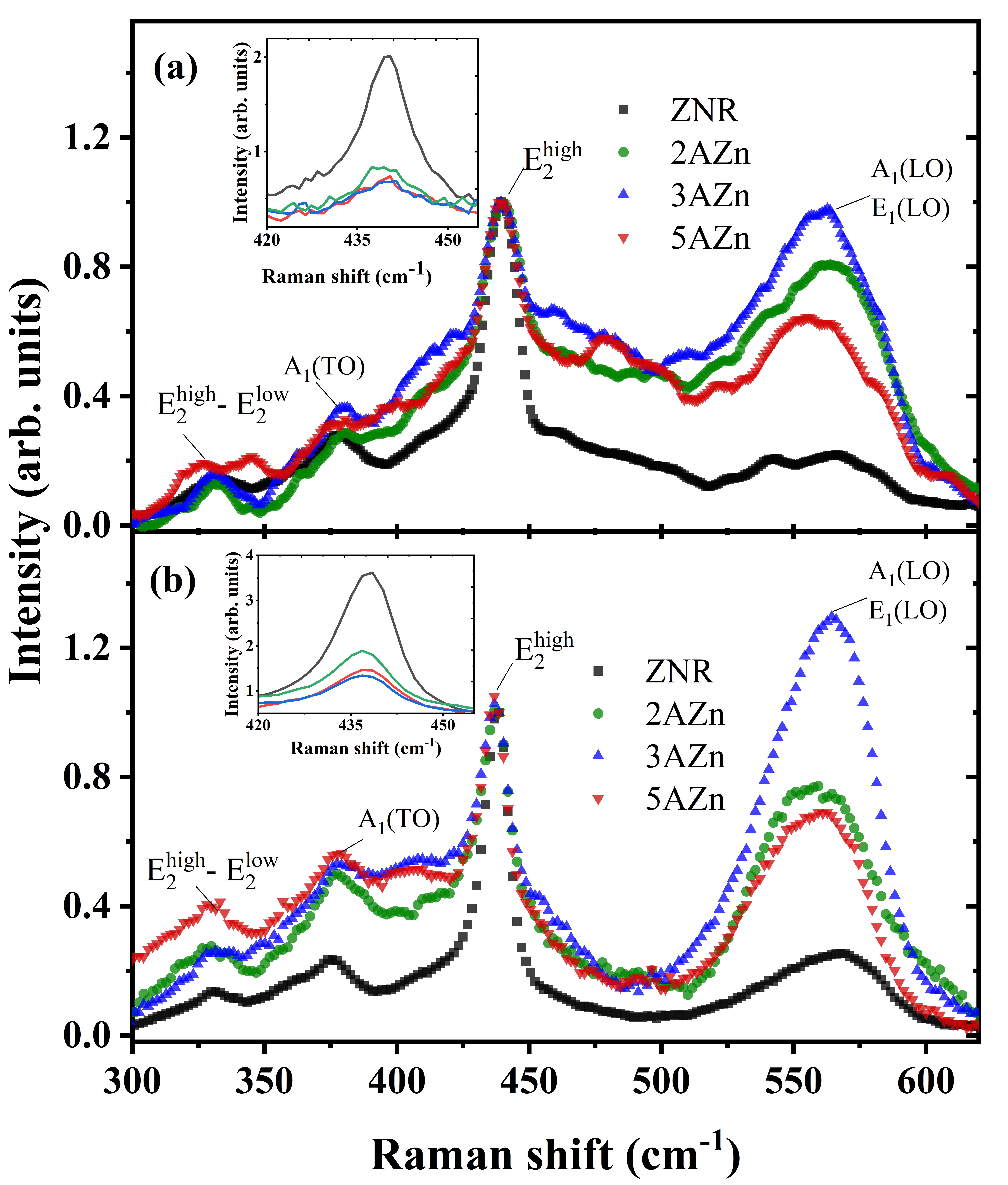}
		\caption{Raman spectra recorded with (a) 638 nm excitation and (b) 532 nm excitation for various samples. All spectra are normalized with the $E_{2}^{high}$ peak intensity at 439 cm$^{-1}$. Inset of (a) and (b) shows evolution of as recorded $E_{2}^{high}$ peak intensity across the samples.}
		\label{Fig:RAMAN-638}
	\end{figure}

The Raman spectra of selected samples  are shown in Fig.\ref{Fig:RAMAN-638}, under 638 nm and 532 nm excitations. The in-plane motion of the O sub-lattice is indicated via the non-polar $E_{2}^{high}$ mode at 439 $cm^{-1}$.  Incorporation of Au NPs significantly compromises the intensity of the $E_{2}^{high}$ mode, as evidenced in the insets of Fig. \ref{Fig:RAMAN-638}. The Zinc sub-lattice vibration counterpart $E_{2}^{low}$ appears at around 99 $cm^{-1}$ (not shown).
	\begin{figure}[t]
	\includegraphics[width=0.83\linewidth]{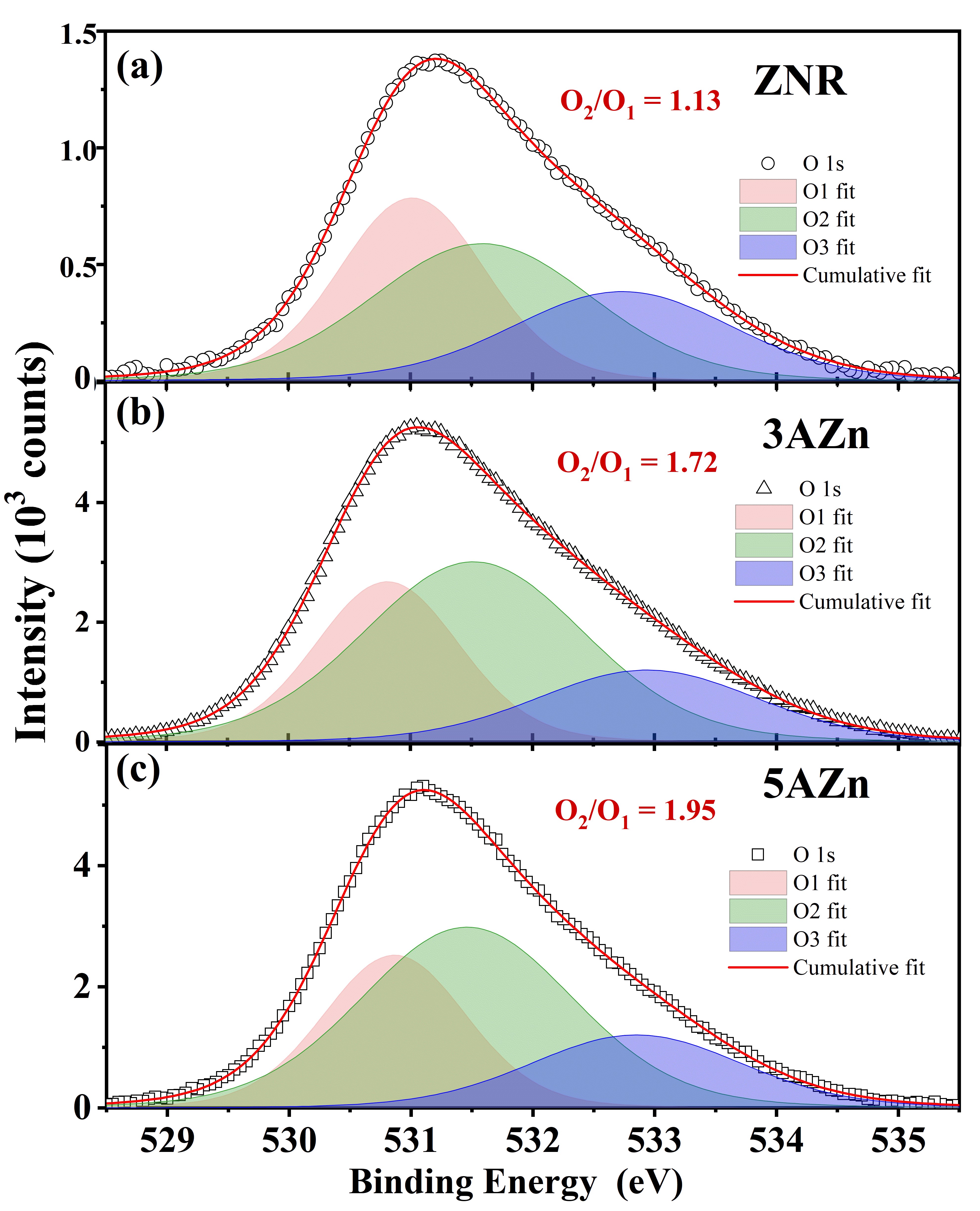}
	\caption{XPS O$^{2-}$ 1s spectra and its deconvoluted into sub-peaks labeled O1, O2 and O3 for (a) bare ZNR (b) 3AuZNR and (c) 5AuZNR. Ratios of area under O2/O1 curves for the samples are recorded.}
	\label{fig:xps}
\end{figure}
\begin{figure}[b]
	\includegraphics[width=0.8\linewidth]{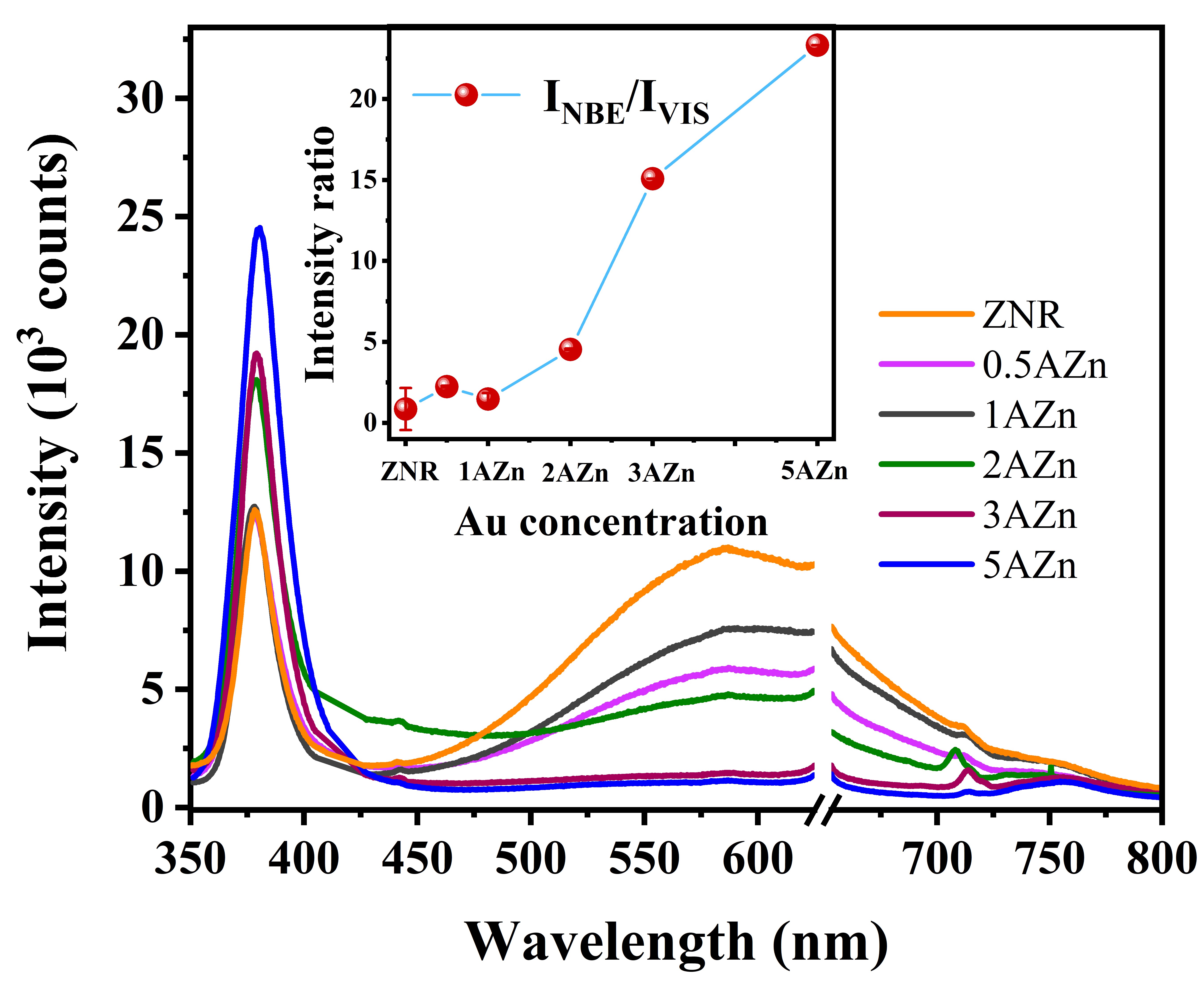}
	\caption{Photoluminescence spectra of bare ZnO nanorods and those with embedded Au NPs. Inset shows the intensity ratio between NBE and peak visible emission intensity with increasing Au NP concentration.}
	\label{fig:PL}
\end{figure}
The wurtzite structure also hosts the  $A_{1}$ and $E_{1}$ modes, split into $TO$ and $LO$ branches. The broad peak centered at 560 $cm^{-1}$ is understood to be a convolution of the $A_1(LO)$ and $E_1(LO)$ modes along with contribution from Fr\"ohlich optical phonon modes \cite{RN700}. These polar modes are susceptible to the local electronic structure and thus sensitive to disorder via defect-induced doping. In the present samples, the feature around 560$cm^{-1}$ increases in prominence with higher inclusion of Au NPs, i.e. samples 2AZn and 3AZn, due to increasing density of defects but then diminishes for the 5AZn sample. The disordered nature of the 5AZn sample with compromised $c-axis$ growth leads to the breakdown of long range Fr\"ohlich interaction\cite{RN701}, adversely affecting the intensity. Overall, modal intensity, in the 400 - 600 $cm^{-1}$ range, benefits from the presence of zone edge phonon modes that increase the phonon density of states \cite{PhysRevB.69.094306}. An increase in intensity with NP inclusion is accompanied by increase in peak width that originates from defect-mediated relaxation of Raman selection rules \cite{Scepanovic2010}. 
The above understanding is further supported by the reproducibility of spectra recorded under both 638 nm and 532 nm excitation and across multiple samples. Under 532 nm excitation, the spectra in Fig. \ref{Fig:RAMAN-638}b benefit from localized surface plasmon-induced enhancement, especially the mixed mode peak centered at 560 $cm^{-1}$. SM table S2 lists the wavenumbers for the various Raman modes of ZnO. 
\begin{figure*}
	\centering
	\includegraphics[width=0.8\linewidth]{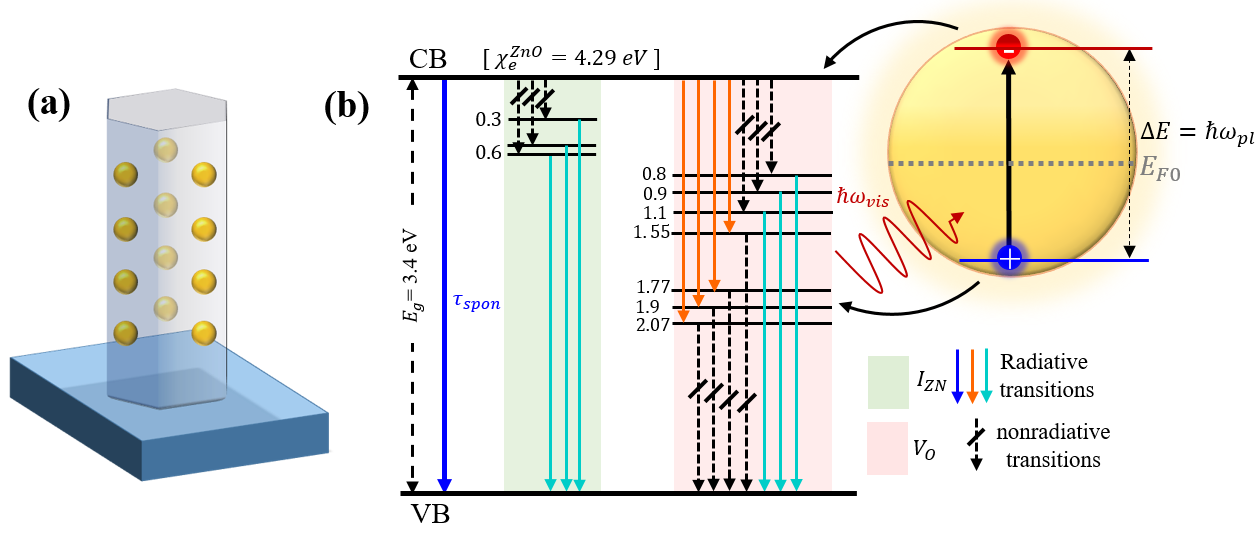}
	\caption{(a) Model of a ZNR embedded with the AuNPs used in FEM calculations.
		(b) ({\it left}) Band diagram of ZnO with defect states and corresponding energies in eV as difference with the CBM. ({\it right}) Schematic showing Au hot carrier ($e^{-}$ and $h^{+}$) generation in energy range $E_{F0}\pm \hbar\omega_{pl}$  and their transfer to ZnO.}
	\label{fig:schematic}
\end{figure*} 

The relative abundance of defects and their evolution across the samples is obtained from XPS analysis. XPS spectra of the oxygen 1s state from bare ZNR and selected AuZNR samples  are shown in Fig.\ref{fig:xps} (a)-(c). Each spectrum is deconvoluted to yield three peaks centred at 529 eV (O1), 530 eV (O2), and 532 eV (O3). While the O3 peak is understood to originate from surface contamination \cite{Kayaci2014}, the O1 and O2 correspond to binding energies of 1s electrons in O$^{2-}$ atoms in stoichiometrically pure and oxygen deficient regions of the lattice, respectively. The ratio of areas under the O2/O1 curves systematically increases across the  samples and almost doubles for the 5AZn sample compared to bare ZNR, quantifying the monotonic increase in $V_O$ abundance upon inclusion of Au NPs.

Inclusion of Au NPs in the ZNR matrix leads to changes in its morphology, crystal structure and stoichiometry, primarily nucleating higher density of $V_O$ and $I_{Zn}$ defects reflected in the PL spectra.  
Selected PL spectra (Fig.\ref{fig:PL}) show that the NBE peak intensity progressively enhances with increasing AuNPs incorporation while the visible emission intensity decreases. The inset shows the change in intensity ratio $I_{NBE}/I_{VIS}$ that enhances with increasing Au NP inclusion, and by more than $\times$20 for the 5AuZNR sample compared to the bare ZNR.\cite{Prajapati2020d,PhysRevB.82.073304} 
Note that the enhancement varies across samples and these numbers are representative. 
The presence of the plasmonically active Au NPs strongly affects the spontaneous decay rates and, thus, luminescence from ZnO. For such low $Q$ plasmonic resonances, the effect is spread over the 500 - 600 nm window, i.e. around the localized Surface Plasmon Resonance(SPR) wavelength of the $\sim$ 30 nm Au NP in ZnO matrix \cite{Prajapati2020d}. 
The physics of the $e-h$ excitation and photon emission under super band gap optical excitation are investigated via a 3D model of the system using FEM simulation in COMSOL Multiphysics using semi-classical formalism, combining the Semiconductor and Wave-Optics module of the software. Fig. \ref{fig:schematic}a shows the in-scale model of ZnO nanorod embedded with Au NPs. 

\begin{figure}
		\centering
		\includegraphics[width=\linewidth]{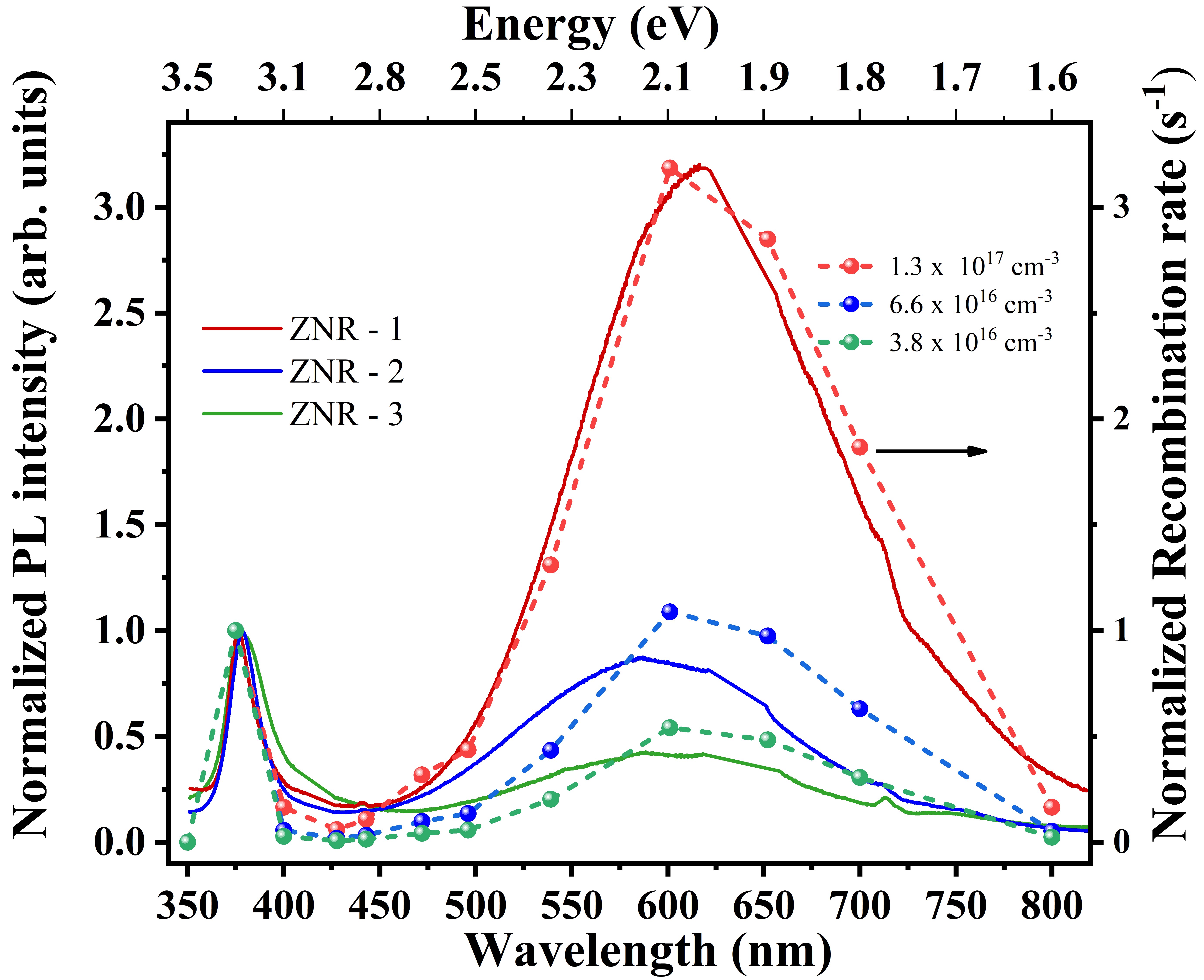}
		\caption{Simulated electron hole recombination rates for the various densities of defects $N_{D}$. The selected experimental PL spectra for the bare ZNR showing three dissimilar defect level emissions guides the selection of corresponding $N_{D}$ in the simulated data. All data are normalized at the NBE emission wavelength(375 nm).}
		\label{fig:ZNR_recombination_rate}
	\end{figure}

The PL spectrum of a bare $n$-type ZNR, i.e. without any Au NP inclusion, is modeled with ten sub-band gap defect states, as shown in Fig. \ref{fig:schematic}b. Each defect state is specified by its energy $E_D$ below the conduction band(CB), number density $N_D$, and the electron and hole capture cross-sections $\sigma_n$ and $\sigma_p$, respectively. Together they determine the electron and hole recombination rates associated with each defect level. 
Defect states with $E_D\leq 0.6 eV$
 correspond to shallow levels arising from variously ionized $I_{Zn}$ and those with $E_D>0.6 eV $ correspond to oxygen vacancy ($V^*_O,V^+_O, V^{++}_O $) states, with the $V^{++}_O $ states having the lowest energies.    
Under continuous wave super band gap energy excitation ($\lambda_{ex} = $ 320 nm), the model calculates the rate of stimulated absorption ($e-h$ pair generation between the VB and CB). Assuming (i) a steady state is achieved between excitation and decay of $e-h$ pairs and (ii) that the Fermi-Dirac statistics adequately describe the distribution of electrons and holes  across the energy states, albeit with different quasi-Fermi energies $E_{Fn}$ and $E_{Fp}$, for electrons and holes respectively. 
Simulating the de-excitation process involves the calculation of various transition rates, including the spontaneous decay rate $R_{spon}$ for direct CB  to VB de-excitation that gives rise to NBE emission given by,
\begin{equation}
	R_{\text {spon }}=\int_{E_{g}}^{\infty} \frac{1}{\tau_{s p o n}} f_{c}\left(1-f_{v}\right) \sqrt{E-E_{g}}  d E
	\label{rspon}
\end{equation}
where $f_{c}$ and $f_{v}$  are the CB and VB electron occupation probabilities and $\tau_{spon}$ (= 2 ns) is the spontaneous emission lifetime for ZnO \cite{guidelli2015enhanced,GALDAMEZMARTINEZ2022100334}.  Similarly, rates for other transitions (Fig. \ref{fig:schematic}b) from the CB edge to individual defect states ($e$-capture) and from defect states to the VB edge ($h$-capture) are calculated. We designate all transitions with emission energy ($\hbar\omega_{em}$) within the PL detection window (3.1 - 1.2 eV) as radiative transitions at the corresponding emission wavelength $\lambda_{em}$, and all other transitions as non-radiative. 
	A plot of the transition rates vs.  $\lambda_{em}$ generates the simulated PL spectrum as shown in Fig. \ref{fig:ZNR_recombination_rate}. The defect states at various $E_D$ are all assumed to have the same $N_D$ but different  $\sigma_n$ and $\sigma_p$. Typically, $\sigma_{n/p}$ would be determined by the selection rules of transition between the states and show wide variation due to variability in local band bending and degree of ionization of the defect state. Broadly, $\sigma_{n/p}$ would decrease with increasing defect state energy difference with the conduction or valence band involved in the radiative transition (Fig. S1 )\cite{10.1063/1.4919100}.  

	Here, variation of $N_D$ alone controls the intensity of visible to NBE emission $I_{NBE}/I_{VIS}$ and is used to mimic the experimentally observed variation across samples (Fig. \ref{fig:ZNR_recombination_rate}), as shown by the blue spectra in Fig. \ref{fig:PL_ALL}. We proceed with $N_D=6.6 \times 10^{16}$/cc that mimics PL of samples showing $I_{NBE}/I_{VIS}\sim 1$. Details of the simulation, including model geometry and material parameters used are available in SM Section S3. 

For the hybrid system, inclusion of Au NPs in ZNRs and UV photoexcitation induce systemic changes that include 
(i) increase in overall defect density $N_D$ in ZNRs due to increased disorder, 
(ii) creation of  Schottky barriers at the Au-ZnO boundaries,  
(iii) absorption of ZnO visible PL in the near field by the NPs to generate LSPs, 
(iv) LSP decay induced hot carrier generation in Au NPs and transfer to ZnO, and finally 
(v) cumulative  modulation in PL spectrum of the hybrid samples. 
  
\begin{figure}
	\centering
	\includegraphics[width=0.8\linewidth]{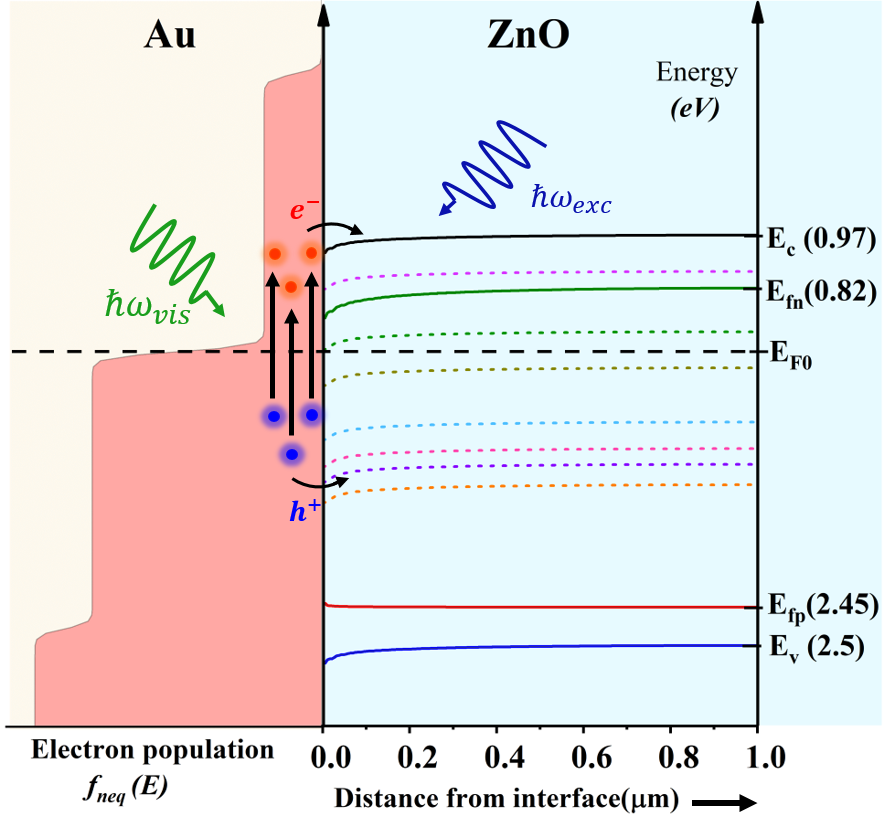}
	\caption{Calculated energy band diagram of the Au-ZnO interface under illumination. (left) Calculated hot electron distribution $f_{neq}$ (Eq.\ref{fneq}) in Au under $\hbar\omega_{vis}$ = 2.3 eV; (right) $E_c$, $E_v$ and quasi-Fermi energies of electrons ($E_{fn}$) and holes ($E_{fp}$) under UV excitation ($\hbar\omega_{exc}$ = 3.8 eV) along with defect states (dotted lines) in ZnO.}
	\label{fig:band_simulation}
\end{figure}

  The PL spectrum from bare ZNRs may be alternatively simulated using a classical dipole emission model by defining a dipole current density ($\mu(\lambda)$) distributed throughout the volume of the ZNRs \cite{Prajapati2020d}. The spectral variation of the dipole moment is given as $\mu(\lambda)\propto \lambda^2\sqrt{I(\lambda)}$, where $I(\lambda)$ is the experimentally recorded PL spectrum of bare ZNR. 
  Beginning with the simulated PL spectrum corresponding to sample ZNR-2 in Fig. \ref{fig:ZNR_recombination_rate}, the PL spectral evolution upon inclusion of Au NPs within the ZNR geometry (Fig.\ref{fig:schematic}(a)) is shown in Fig. S2(SM Section S3). With increasing Au NP density the $I_{NBE}$ changes negligibly and the decrease in $I_{VIS}$ is far muted than that observed experimentally, demonstrating that plasmonic absorption, modeled using classical electrodynamics alone does not account for the observed quenching of visible emission. 
  Within the semi-classical framework, the work function of Au ($\phi^{Au}$), along with the electron affinity of  ZnO ($\chi_e^{ZnO}$) set the energy scale of the hybrid system.  At equilibrium without any photoexcitation, solution of the Poisson equation shows that the CB edge in the $n$-type ZnO bends upwards at the Au-ZnO interface creating a Schottky barrier and local depletion in $n_e$ (SM Fig. S3). However, upon super band gap excitation ($\lambda_{ex} \sim $ 320 nm), the increase in photoexcited carriers induces a nonequilibrium steady state in ZnO, increasing $n_e$ with the CB edge bending down by 0.15 eV below its flat band energy, as shown in Fig. \ref{fig:band_simulation}  (see SM Section S4 for further details). 
The visible component ($\hbar\omega_{vis}$) of the ensuing PL (Fig. \ref{fig:ZNR_recombination_rate}), with strong overlap with the absorption of the NPs, excites LSPs of energy $\hbar\omega_{pl} (=\hbar\omega_{vis})$. Once excited, the  LSPs dephase in the time scale $\tau_1\sim$ 10 fs \cite{Brongersma2015,Voisin2001} and decay radiatively and non-radiatively, depending on NP size \cite{Mertens2017a}. In smaller NPs, the latter dominates causing $e-h$ pair generation and subsequent thermalization with $\tau_2\sim$ 0.1 - 100s of ps \cite{Furube2007}. 
Under continuous excitation, hot carriers in the NPs reach a steady state, non-equilibrium distribution $f_{neq}(E)$, that departs significantly from the equilibrium distribution $f_{eq}(E)$ in the energy range   $E_{F0}\pm \hbar\omega_{pl}$, where $E_{F0}$ is the Fermi energy prior to photoexcitation \cite{Bernardi2015b, Brown2016, Besteiro2017}.
For spherical NPs, $f_{neq}(E)$ may be determined from Eq. \ref{fneq}, following Kornbluth \textit{et al.} \cite{Kornbluth2013a};  
\begin{equation}
	\begin{split}
		f_{neq}(E)= f_{eq}(E) + \frac{N_{A} \tau}{\rho\hspace{0.03cm}\hbar\omega_{pl} } \tanh \left(\frac{\beta}{2}(E-E_{F0})\right) \\ \times  \   \frac{1}{1+2e^{-\beta\hbar \omega_{pl} }\cosh \left(\beta(E-E_{F0})\right)}
		\label{fneq}
	\end{split}
\end{equation}
where $N_{A}$ is the number density of  plasmons absorbed per unit time, $\tau (< \tau_2)$ is a relaxation time constant, $\rho(E)$ is the density of states, $\beta = 1/k_BT$ and $\hbar\omega_{pl} =\hbar\omega_{vis}$.  The left panel of Fig. \ref{fig:band_simulation} plots  $f_{neq}$ calculated for $E_{F0}$ = 5.1 eV at 300 K and $\hbar\omega_{vis}$ = 2.3 eV(see SM Section S5 for further details).
$N_A$ is determined by several parameters including the number of embedded Au NPs, intensity of visible emission from ZnO, which is dependent on the intensity of the UV excitation ($\hbar\omega_{exc}$).  Finally the quantum efficiency of hot carrier generation and the fraction transferred to ZnO, aided by the downward band bending at the interface determine $N_A$. SM Fig. S5 plots the $f_{neq}$ for various values of $N_A$ and $\hbar\omega_{pl}$. 
Hot electrons transferred to the ZnO CB would enhance NBE emission, which is incorporated in the present model by recalculating  $R_{sp}$ (Eq. \ref{rspon}) by modifying the ZnO CB occupation by addition of $f_{neq}$ distribution of electrons. The increased rate $R'_{sp}$ is given as:
\begin{equation}
	R'_{sp} =\int_{E_{g}}^{\infty} \frac{1}{\tau_{sp}} \left( f_{c}+f_{neq}\right) \left(1-f_{v}\right) \sqrt{E-E_{g}}  d E
\end{equation}
 \begin{figure}[t]
	\centering
	\includegraphics[width=0.9\linewidth]{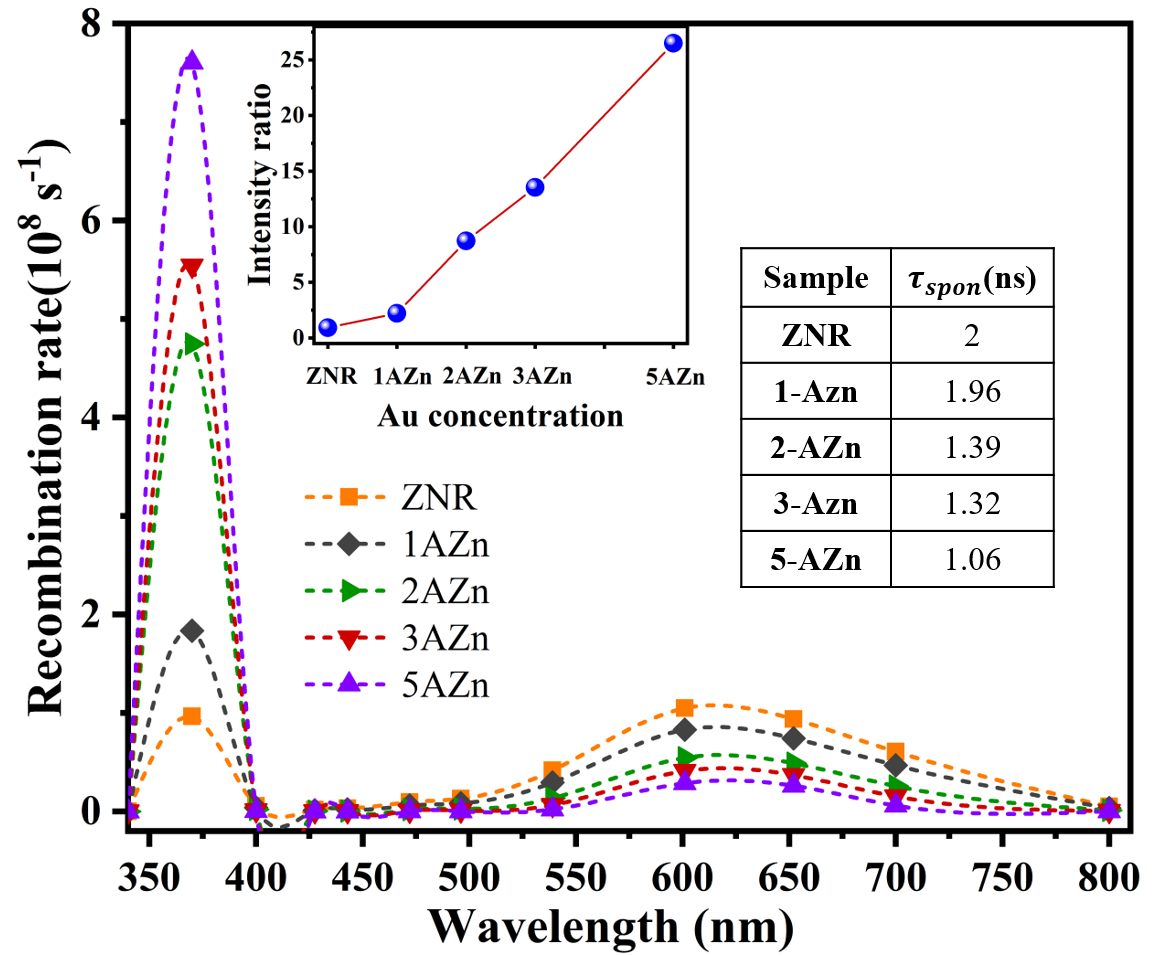}
	\caption{Spectral variation of calculated recombination rate for ZNRs with increasing concentration of Au NPs}
	\label{fig:rate}
\end{figure}
Increasing Au NP density across samples is reflected in $f_{neq}$ via increasing  $N_{A}$ in Eq. \ref{fneq}, which is guided by the experimentally detected NBE emission enhancement.  
The increase in $R_{sp}$ is reflected in time resolved PL measurements that show a decrease in NBE spontaneous emission life-time as summarized in SM table S6 in the presence of plasmonic NPs \cite{wang2021enhanced,RN713,RN714,RN715,RN716}. Recalculating the transition rates with $R'_{sp}$ and thus $\tau_{sp}$ thus incorporates the hot electron contribution from NP inclusion, which intensifies NBE emission and provides a direct coupling that quantifies the cross-talk between the components. The nonequilibrium steady state reached under CW photoexcitation also modifies the  probability of  electron occupation across the defect states. Though $N_D$ is assumed to be the same across the defect states, the number changes dynamically under photoexcitation, affecting the kinetics of the PL process. SM Fig. S4 plots the occupancy of the states before and after hot electron transfer showing that DL state occupancy $\sim$ 1, with two consequences. Firstly, the ionized DL defect states ($V_O^+, V_O^{++}$) under high occupancy cease to exist at the designated energies, selectively reducing their individual $N_D$. The abundance and occupation of defect states is non-trivial to estimate, especially with photoexcitation. The simulation uses Shockley-Read-Hall statistics to calculate the probability of occupation of the defect states \cite{rhoderick1988metal}. At equilibrium the fraction of any ionized donor species is given by $N^+_D/N_D \sim 1/(1+2\exp(E_{F0}-E_D)/kT)$ \cite{book}. 
	\begin{figure}
	\centering
	\includegraphics[width=0.9\linewidth]{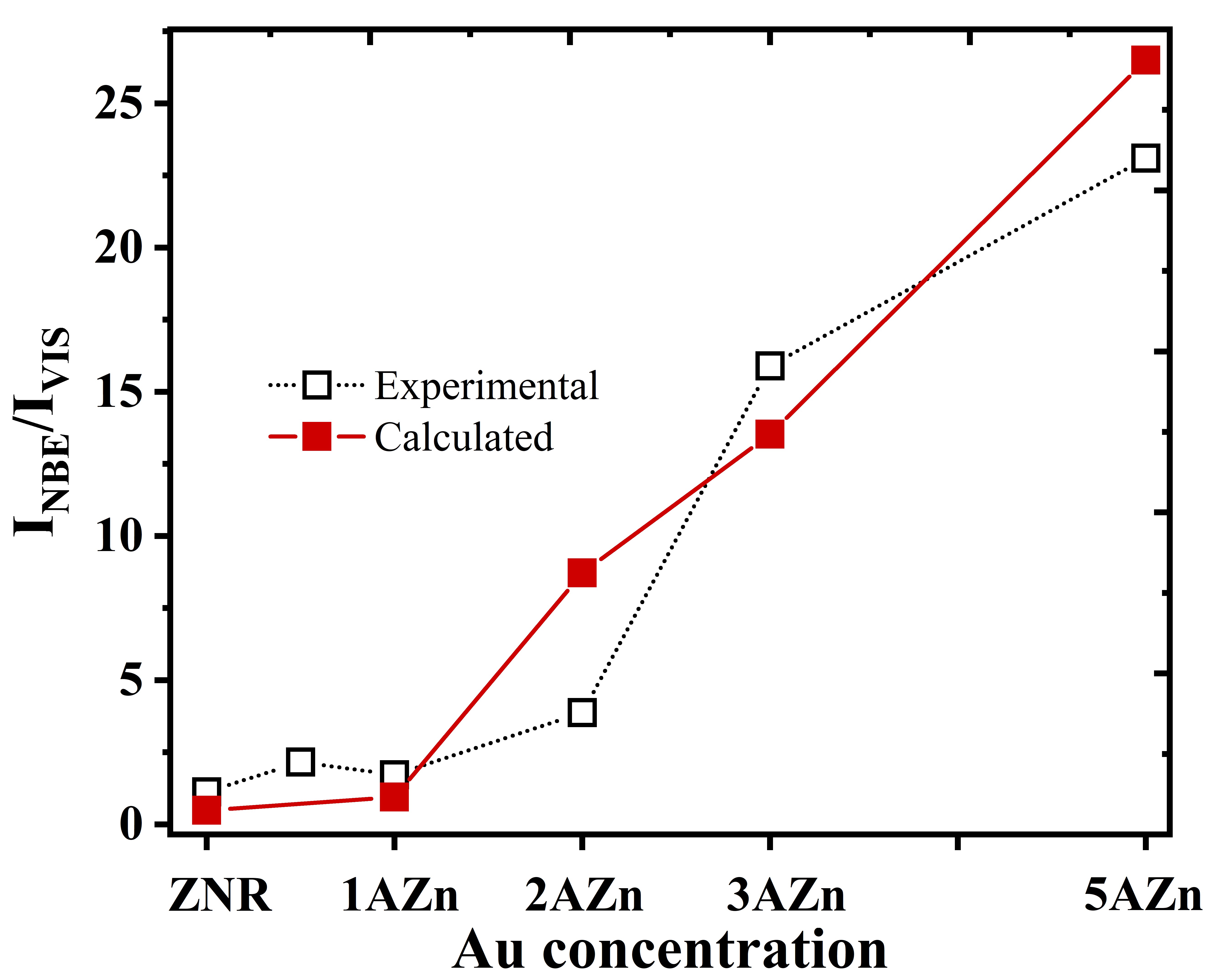}
	\caption{Calculated relative intensity ratio between NBE and visible emission with increasing Au NP concentration along with the calculated recombination rates.}
	\label{fig:calculation}
\end{figure}
Under photoexcitation, $E_{Fn}$ lies closer to $E_C$, decreasing $N^+_D/N_D$ and quenching the relative abundance of the DL ionized defect states, as discussed above. 
Secondly, the energy proximity of the DL states with $E_{F0}$ indicates presence of non-radiative charge transfer pathways between Au and ZnO. Hot holes may diffuse and recombine with DL state electrons or vice-versa, thereby reducing the availability of DL electrons, compromising both NBE and visible emission \cite{guidelli2015enhanced}. Under CW illumination, these competing processes would reach a steady state, compromising radiative transitions to and from the DL states in ZnO and thus, visible emission.   
The condition is mimicked in the simulation by a decrease in $N_D$ preferentially for the states with occupancy close to 1. Beginning with model parameters for bare ZNR that yield comparable maximum recombination rates for NBE and visible emission, Fig.\ref{fig:rate} plots the evolution with increasing Au NP inclusion. As $N_A$ progressively increases, raising $R'_{sp}$, decrease in $N_D$s for selected DL states based on calculated occupancy of the states decrease visible emission. The table in the inset of Fig.\ref{fig:rate} lists the calculated decrease in $\tau_{sp}$ with the changing parameters. 
The efficacy of the simulations and optimal choice of  parameters in modeling experimental results are further shown in Fig.\ref{fig:calculation}, which plots the $I_{NBE}/I_{VIS}$ ratio for samples with increasing Au NP density across the experimental and simulated results. Agreement between the results evidence the capability model in capturing the key ingredients of light-matter interactions discussed here.

Though the experimental results modeled here follow previous reports \cite{RN713, RN714, RN715, RN716, khan2019enhanced, wang2021enhanced, guidelli2015enhanced} the near complete quenching of visible emission has not been reported earlier. Size and density of the plasmonic inclusions play a crucial role in determining their effect on the PL of the semiconductor.  Increasing NP density above a threshold leads to quenching of PL due to charge transfer from ZnO to Au NPs \cite{chiu2018plasmon}, which is controlled via both geometric parameters and the nature of band alignment across the components \cite{Prajapati2020d}.  Further, increase in UV excitation ($\hbar\omega_{exc}$) intensity enhances PL NBE emission more than visible emission in plasmonic hybrids, in contrast to bare ZnO where both components are seen to increase in tandem \cite{guidelli2015enhanced}.  
In the model, intensity increase if effected via increasing $N_A$, which again selectively enhances the NBE emission more than the visible. 
	
\section{Conclusion} 
This investigation models the photoluminescence of an Au - ZnO hybrid plasmonic system using a semiclassical model, employing finite element method calculations. Experimental results evidence complete quenching of visible emission and simultaneous increase in NBE emission intensity, with increasing Au NP inclusion in ZnO nanorods. The result is non-trivial since inclusion of Au NPs substantially increases morphological disorder in the ZnO lattice and increases oxygen defect density as investigated by a combination of structural and spectroscopic probes. The model, incorporating ZnO's band properties with sub-bandgap defect states, reproduces the photoluminescence from bare ZnO nanorods and its variation across samples. The effect of plasmonic NPs under UV excitation is incorporated via calculating localized surface plasmon-mediated hot electron generation in Au and subsequent transfer to the conduction band of ZnO, thus enhancing NBE emission. Finally, reduction in visible emission is partially attributed to near-field absorption by the Au NPs and partially to passivation of the deep-level ionized defect states due to high electron occupancy under photoexcitation and non-radiative charge transfer to Au NPs. Overall, the model successfully combines the photo-stimulated excitation and de-excitation processes in a wide band gap semiconductor with plasmon-mediated hot electron generation in metal nanoparticles to simulate the evolution of photoluminescence spectra of the semiconductor. With increasing interest in hybrid plasmonic systems and novel applications, the results offer a modeling platform for engineering their optoelectronic and photocatalytic properties.    
\hspace{2cm}
\begin{acknowledgments}
The authors thank Harikrishnan G. (IISER TVM) for valuable inputs on XPS analysis. The authors acknowledge ISTEM, Govt. of India for access to COMSOL Multiphysics software and financial support from SERB, Govt. of India (CRG/2019/004965). KS and SC acknowledge PhD fellowship from IISER TVM.
\end{acknowledgments}

 KS and SC contributed equally to this work. 

\bibliography{AuZnOref}

\end{document}


\title[Supplemental Material]{ Supplemental Material\\
	\Large{Leveraging Plasmonic Hot Electrons to Quench Defect Emission in Metal - Semiconductor Nanostructured Hybrids: Experiment and Modeling}}

\author{Kritika Sharu}
\email{kritika.iiser16@iisertvm.ac.in}
\affiliation{School of Physics, Indian Institute of Science Education and Research Thiruvananthapuram, Kerala 695551, India}

\author{Shashwata Chattopadhyay }
\affiliation{School of Physics, Indian Institute of Science Education and Research Thiruvananthapuram, Kerala 695551, India}

\author{K. N. Prajapati}
\affiliation{School of Physics, Indian Institute of Science Education and Research Thiruvananthapuram, Kerala 695551, India}

\author{J. Mitra }
\email{j.mitra@iisertvm.ac.in}
\affiliation{School of Physics, Indian Institute of Science Education and Research Thiruvananthapuram, Kerala 695551, India}
\maketitle

\vspace{-1cm}
\section*{S1. Sample Preparation}
The bare ZNR and Au-ZNR samples were grown hydrothermally on   Zn foil ($0.25$ mm thick; Alfa Aesar).  Firstly, a Zn foil was cut to $\sim$ 5 square mm dimension and cleaned by subsequent sonication in acetone, ethanol and DI water, followed by 2 minutes of ozone treatment.  For growing bare ZNRs, the Zn foil was kept floating on the surface of equimolar (25 mM) aqueous solution of zinc nitrate hexahydrate ($Zn(NO_{3})_{2}.6H_{2}O$) and
hexamethylenetetramine (HMTA, $C_{6}H_{12}N_{4}$) at 80$\degree$C for 9 hours.  For fabrication of Au embedded ZNRs, the as purchased AuNP solution (in citrate buffer) with density $\sim 10^{11}$ nanoparticles per cc was added to the precursor solution in selected volume, keeping other growth parameters unchanged. To increase the Au:ZnO atomic fraction, the AuNP suspension volume was increased and the zinc nitrate hexahydrate and HMTA stock solution volume was decreased simultaneously. The volumes of stock solution used for the growth of samples with least NP inclusion (bare ZNR)
to maximum (5AZn) NP inclusion are detailed in table \ref{tab:concentration}. 
The prepared samples were rinsed with DI water and dried in an oven at 65$\degree$C for few hours before characterization.

\begin{table}[!h]
	\begin{tabular}{ | c | c | c | c | }
		
		\hline 
		Sample  & \makecell[c]{AuNP solution ($10^{11}$NPs/ml)\\ volume (ml)}  &\makecell[c]{$Zn( NO_{3})_{2} .6H_2O$+ HMTA \\ solution volume (ml)}& \makecell[c]{Au:ZnO atomic ratio \\in growth solution}\\
		\hline
		\textbf{ZNR}  & 0.0  & 100   & 0  \\
		\textbf{0.5AZn}
		&
		{\fontfamily{ptm}\selectfont 0.5}
		&  {\fontfamily{ptm}\selectfont 64 }  & $\displaystyle 4.36\ \times \ 10^{-5}$ \\
		\textbf{	 1AZn}
		& 
		1
		
		& {\fontfamily{ptm}\selectfont 64} & $\displaystyle 8.73\times \ 10^{-5}$ \\

		\textbf{2AZn}
		
		& 
		{\fontfamily{ptm}\selectfont 2}
		
		& {\fontfamily{ptm}\selectfont 48} & $\displaystyle 2.37\times \ 10^{-4}$ \\

		\textbf{3AZn}
		
		&
		{\fontfamily{ptm}\selectfont 3}
		
		& {\fontfamily{ptm}\selectfont 32}  & $\displaystyle 5.41\times \ 10^{-4}$ \\

		\textbf{5AZn}
		
		&
		{\fontfamily{ptm}\selectfont 5}
		
		& {\fontfamily{ptm}\selectfont 16}  & $\displaystyle 1.8\times \ 10^{-3}$ \\
		\hline
	\end{tabular}
	\caption{Au-ZnO atomic fraction in growth solution for various ZnO nanorod samples}
	\label{tab:concentration}
\end{table}

\section*{S2. Raman spectra }

\begin{table}[h]
	\centering
		\begin{tabular}{ | l | c | c | c |} 
			\hline
			\multirow{2}{*}{Mode} & \multicolumn{3}{l|}{\hspace{1.5cm} Raman shift ($cm^{-1}$) } \\
			\cline{2-4} 
			& 	This work  & Ref. \cite{RN700}  & \cite{Scepanovic2010}  \\
			\hline
			$\textbf{E}_{\textbf{2}}^{\textbf{low}}$    & 100 & 99  & 99  \\ 
	\hline
			$\textbf{E}_{\textbf{2}}^{\textbf{high}}$-$\textbf{E}_{\textbf{2}}^{\textbf{low}}$    & 331 & 333  & 331 \\
	\hline
				$\textbf{A}_{\textbf{1}}\textbf{(TO)}$   & 380 &  380 & 384  \\
	\hline
			$\textbf{E}_{\textbf{2}}^{\textbf{high}}$   & 440  & 438 & 439  \\
		\hline
			\makecell{$\textbf{A}_{\textbf{1}}\textbf{(LO)}$, $\textbf{E}_{\textbf{1}}\textbf{(LO)}$\\ mixed modes} & 560  &\makecell{$A_{1}(LO)$ - 574 \\  $E_{1}(LO)$ - 590}  & \makecell{$A_{1}(LO)$ - 569 \\  $E_{1}(LO)$ - 586}   \\
			\hline
 		\end{tabular}
		
		\caption{Raman modes observed in ZNR samples and their assignment. Comparison  with reports on  ZnO single cyrstal \cite{RN700} and ZnO nanopowder \cite{Scepanovic2010} are included.}
		\label{tab:raman}
\end{table}
\section*{S3. Finite element method modeling of PL spectra of ZnO}

\begin{figure}[h]
	\centering
	\includegraphics[width=0.5\linewidth]{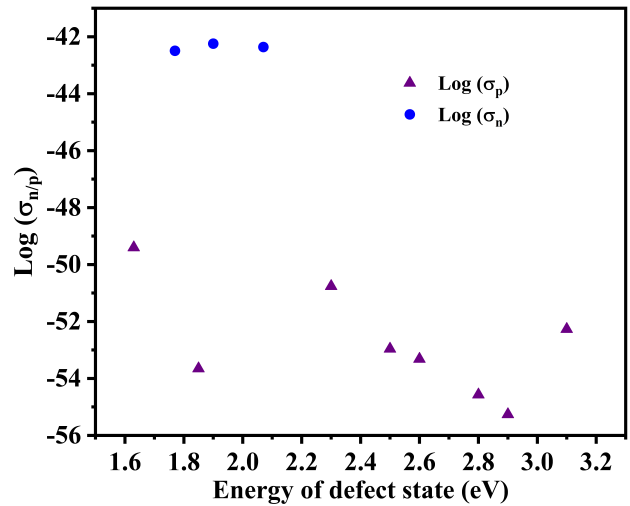}
	\caption{ Variation ln($\sigma_p$) and ln($\sigma_n$) with energy difference of defect states from VB and  CB respectively.}
	\label{Fig:4}
\end{figure}

FEM solver COMSOL Multiphysics 5.3(a) with Wave Optics and Semiconductor modules was used  to simulate the photoluminescence spectra of Au embedded ZnO nanorods. The maximum and minimum mesh grid size was optimized to be $13 nm$ and $2 nm$. The material properties for ZnO used in the model as given in Table \ref{tab:Material properties used in simulation } \cite{10.1063/1.1331089}.
\begin{table}[h]
	\centering
\begin{tabular}{|l | l |}
		\hline
		Property & Value\\
		\hline
		Relative permittivity & 8.3\\
		Band gap & 3.4 eV\\
		Electron affinity (CBM)& 4.29 eV\\
		Integrated density of states(CB) & $4.8 \times 10^{18} cm^{3}$\\
		Integrated density of states(VB) & $4.8 \times 10^{18} cm^{-3}$\\
		Electron mobility & $200\ cm^{2}V^{-1}s^{-1}$\\
		Hole mobility & $50\ cm^{2}V^{-1}s^{-1}$\\
		Dopant density (n-type) & $1 \times 10^{12} cm^{-3}$\\		\hline
	\end{tabular}
	\caption{ZnO Material properties used in simulation }
	\label{tab:Material properties used in simulation }
\end{table}
\begin{table*}[h]
	\centering
	\begin{tabular}{ | c | c | c | c |} 
		\hline

		Defect state & $\Delta E$ below CBM (eV) & $\sigma_n \times10^{-19} $cm$^2$ & $\sigma_p \times10^{-22} $cm$^2$ \\
		\hline
		$I_{Zn}$ & 0.3 & 0.3 & 20\\
		$I_{Zn}$ & 0.5 & 1 & 100\\
		$I_{Zn}$ & 0.6 & 1 & 700\\ \hline
		$V_O$ & 0.8 & 1.6 & 700\\
		$V_O$ & 0.9 & 2 & 10\\
		$V_O$ & 1.1 & 2.5 & 90\\
		$V_O$ & 1.55 & 0.5 & 500\\
		$V_O$ & 1.77 & 3.5 & 0.035\\
			$V_O$ & 1.9 & 4.5 & 0.05\\
		$V_O$ & 2.07 & 4 & 0.008\\
		\hline
	\end{tabular}
	\caption{Defect state Energy, and electron and hole capture cross sections ($\sigma_n, \sigma_p$). The shallow states are identified as interstitial Zn states ($I_{Zn}$) and the deep level states are identified as variously ionized oxygen vacancy states ($V_O: V_O^*, V_O^+, V_O^{++}$)}
	\label{tab:Properties defect states}
\end{table*}
The model includes sub band gap defect states as given in table \ref{tab:Properties defect states}, as specified via energy difference with the CB edge $\Delta E$ \cite{10.1063/1.4919100}. The $\Delta E$ are chosen such that transitions from the CB to defect states and from defect states to VB reproduce the experimentally observed PL emission in the visible wavelength regime.Figs. \ref{Fig:4} plots the  optimized values of the  hole and electron capture cross sections, ln($\sigma_p$)  and ln($\sigma_n$) of each state and their energy difference with the  CB minima or VB maxima, as applicable. 

\begin{figure}
	\centering
	\includegraphics[width=0.5\linewidth]{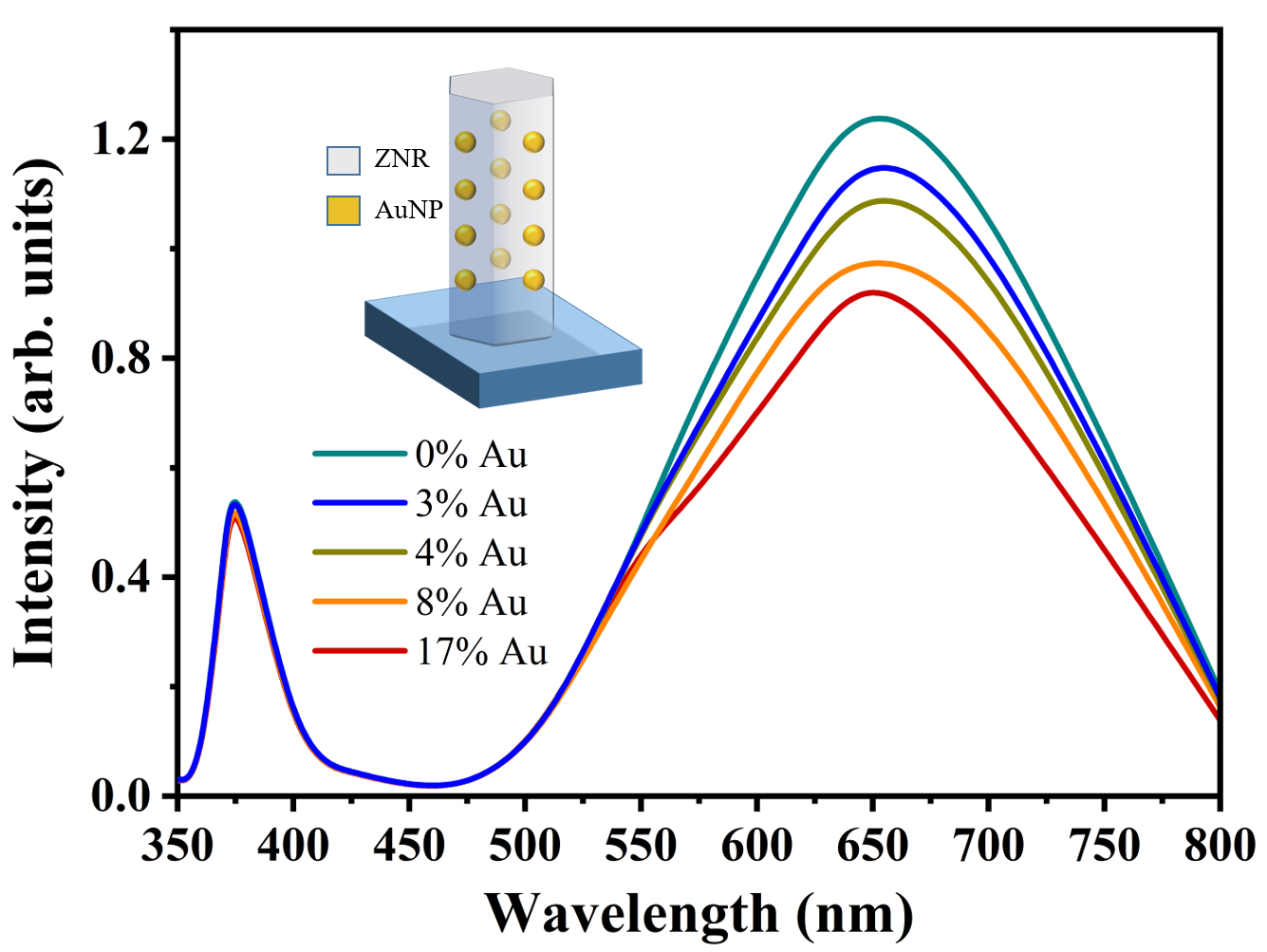}
	\caption{Calculated radiated power from ZNR and AuNP embedded ZNR samples. AuNP inclusion is expressed as volume fraction of the ZNR volume.}
	\label{wo-pl}
\end{figure}

The simulations were performed with an excitation wavelength of 320 nm and initial defect state density  $N_{D}=1\times10^{17} cm^{-3}$ for all defect states. The electron and hole thermal velocity $v_{n}$ and $v_{p}$ are assumed to be $2.0242\times10^{7} cms^{-1}$ and $1.562\times10^{7} cms^{-1}$ respectively \cite{10.1063/1.4919100}. 
Plasmonic absorption peak of 30 nm diameter Au NPs in ZnO environment lies around 575 nm \cite{Prajapati2020d}. Au NPs embedded in ZnO will more effectively absorb PL emission in the visible thus reducing the far field radiation from the ZnO nanorods. To simulate this purely plasmonic absorption of the PL near field, the WaveOptics module of  COMSOL Multiphysics was used. Inset figure \ref{wo-pl} shows the simulated geometry, comprising of a hexagonal ZNR with length = 300 nm and radius 50 nm, with Au NP (diameter = 30 nm) embedded inside. The refractive index of ZnO was taken to be 1.75 + 0.01\textit{i} and the data of Rakic \textit{et al}\cite{rakic1998optical} describes the dielectric function of Au. The entire domain is approximately $1.1\mu m \times0.17\mu m \times0.17 \mu m$ and bounded with perfectly matched layers to absorb the outgoing radiation. A volume current density element is defined over the entire ZNR.  The spectral variation of volume current density may be back calculated from the PL spectrum of a bare ZNR sample. The far field radiation intensity is calculated by integrating the Poynting vector over domain boundary at a distance of 1$\mu$m from the ZNR to avoid any near field scattering effect.
Simulations were performed with increasing number of AuNPs embedded inside the ZNR, expressed as volume fraction change from 0\% (no AuNP inclusion) to 17\% (max. inclusion). Figure \ref{wo-pl} plots the calculated spectra across the samples, showing decrease in visible emission with increasing Au NP inclusion. The reduction of radiated power for visible wavelength regime is attributed to increased plasmonic absorption in this range. However, the reduction in visible emission is not pronounced and moreover the radiated power at the NBE emission wavelength remains unmodified for all AuNP inclusions, demonstrating that plasmonic absorption alone does not account for the observed changes in the PL spectra of ZnO with Au NP inclusion.
\newpage
\section*{S4. Energy band diagram at the ZnO - Au interface}
FEM simulation of the energy band diagram at the  Au - ZnO interface, using the Semiconductor module was performed to comprehend the energetics of electron transfer between the components, especially under photo excitation. 
The material properties of ZnO used are as given in tables S3 and S4r while those for Au is given in Table \ref{tab:Material properties of Au }.
\begin{table}[h]
	\centering
	\begin{tabular}{|l|l|}
		\hline
		Property & Value\\
		\hline
		Work Function & 5.1 eV\\
		Optical properties & Rakic et. al \cite{rakic1998optical} \\
		\hline
	\end{tabular}
	\caption{Material properties of Au. }
	\label{tab:Material properties of Au }
\end{table}

\begin{figure}[h]
	\centering
	\includegraphics[width=0.7\linewidth]{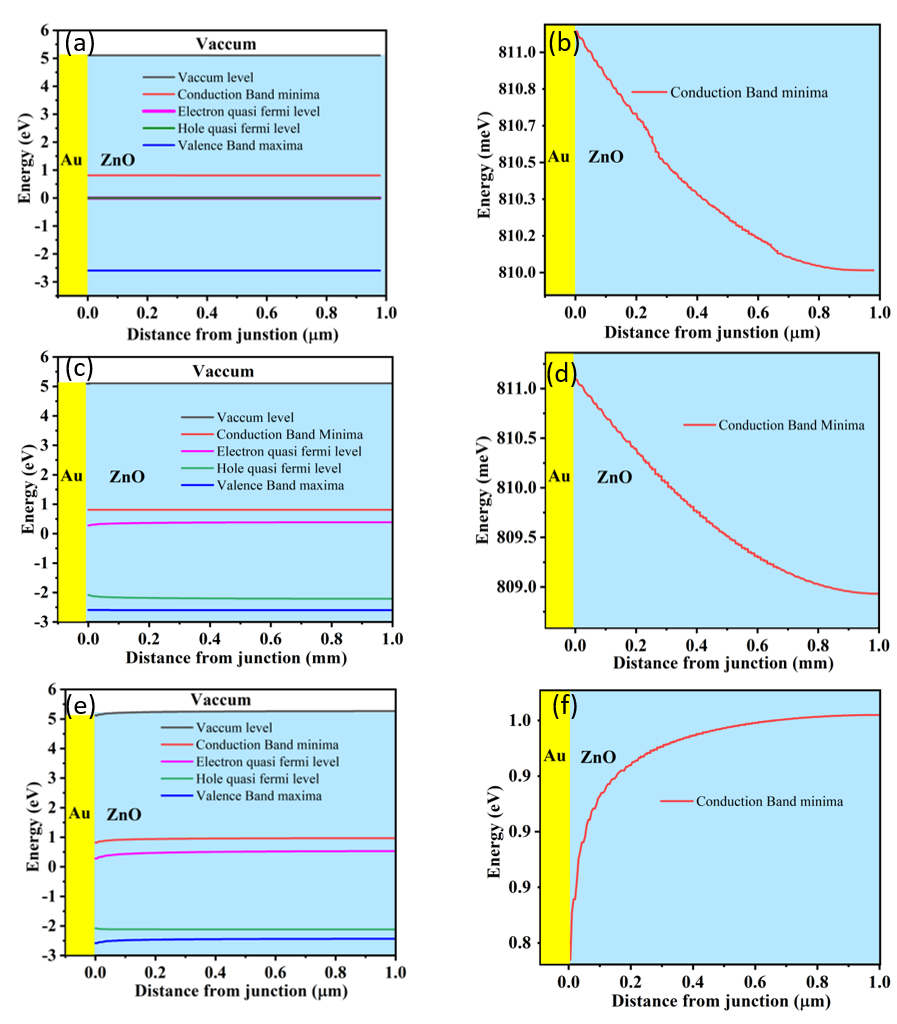}
	\caption{Energy band diagram of simulated system (a) Without defects and photo-excitation. (b) Enlarged CB plot without defects and photo-excitation. (c) Without defects in presence of photo-excitation. (d) Enlarged CB plot without defects in presence of photo-excitation. (e) With inclusion of defects and photo-excitation. (f) Enlarged CB plot with inclusion of defects and photo-excitation.}
	\label{Fig:2}
\end{figure}

All simulations were conducted at T = 300 K and the main results are summarized below;
\begin{enumerate}
	\item For ZnO without any defect states, without optical excitation, at equilibrium:  The CB edge has a upward bent at the junction showing formation of a Schottky barrier. Fig. \ref{Fig:2} b shows a magnified version of the plot in  Fig. \ref{Fig:2} a.
	The $n$-type doping of bare ZnO is evidenced by the $E_{F0}$ (at 0.0 V) lying close to the CB than the VB edge.
	\item For ZnO without any defect states and with optical excitation at 320 nm.  The CB edge has a upward bent at the junction showing formation of a Schottky barrier. Fig. \ref{Fig:2} d shows a magnified version of the plot in  Fig. \ref{Fig:2} c. In the non-equilibrium case the corresponding chemical potential of the electron and hole species are defined by their respective quasi Fermi levels.
	\item In either of the above cases the band bending shows formation of an electron depletion layer in the semiconductor. The extent of CB bending at the interface is slightly modified under optical illumination compared to that calculated  in the dark.
	\item For ZnO with defect states and with optical excitation at 320 nm:  The CB edge has a downward bent at the junction, in the semiconductor. Fig. \ref{Fig:2} f shows a magnified version of the plot in  Fig. \ref{Fig:2} e.
	\item The modification in the band diagram under illumination and in the presence of the defect states show  that the later play a significant role in determining the band alignment at the interface and enable electron transfer from Au to ZnO, under illumination.\\
	\item Comparison of the free electron density under illumination, indicate that it increases by an order in magnitude, from $1\times10^{15} m^{-3}$ to $1\times10^{16} m^{-3}$ at the interface with inclusion of defect states in ZnO. 
\end{enumerate}
 
\begin{figure}[h]
	\centering
	\includegraphics[width=0.5\linewidth]{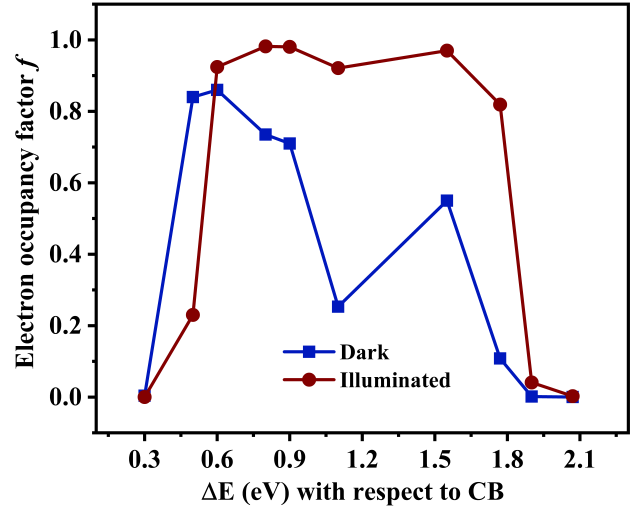}
	\caption{Electron occupancy factor for individual defect states as a function of  their energies below the CB, in the dark and under photoexcitation at 320 nm.}
	\label{Fig:3}
\end{figure}
Figure \ref{Fig:3} shows the calculated electron occupation factor $f$ for the defect states near the Au-ZnO junction with and without optical illumination. It shows that in the dark condition (equilibrium state) shallow level states (above $E_{F0}$), have low electron occupancy, thus are ionized having lost electrons to the CB of ZnO, aiding the $n$-type nature of ZnO.   
The DL states mostly being ionized $V_O$ states intrinsically have lower occupancy. 
However, the occupancy changes under 320 nm illumination, the occupancy of these DL states increase to $\sim 1$, which is commensurate with the increased electron density in ZnO. The ionized defect states with high occupancy will cease to exist at the assumed energies which decrease their relative abundance at the cost of the of unionized defect states.

\section*{S5. Plasmon induced hot carrier generation}
\begin{figure}[h]
	\centering
	\includegraphics[width=0.9\linewidth]{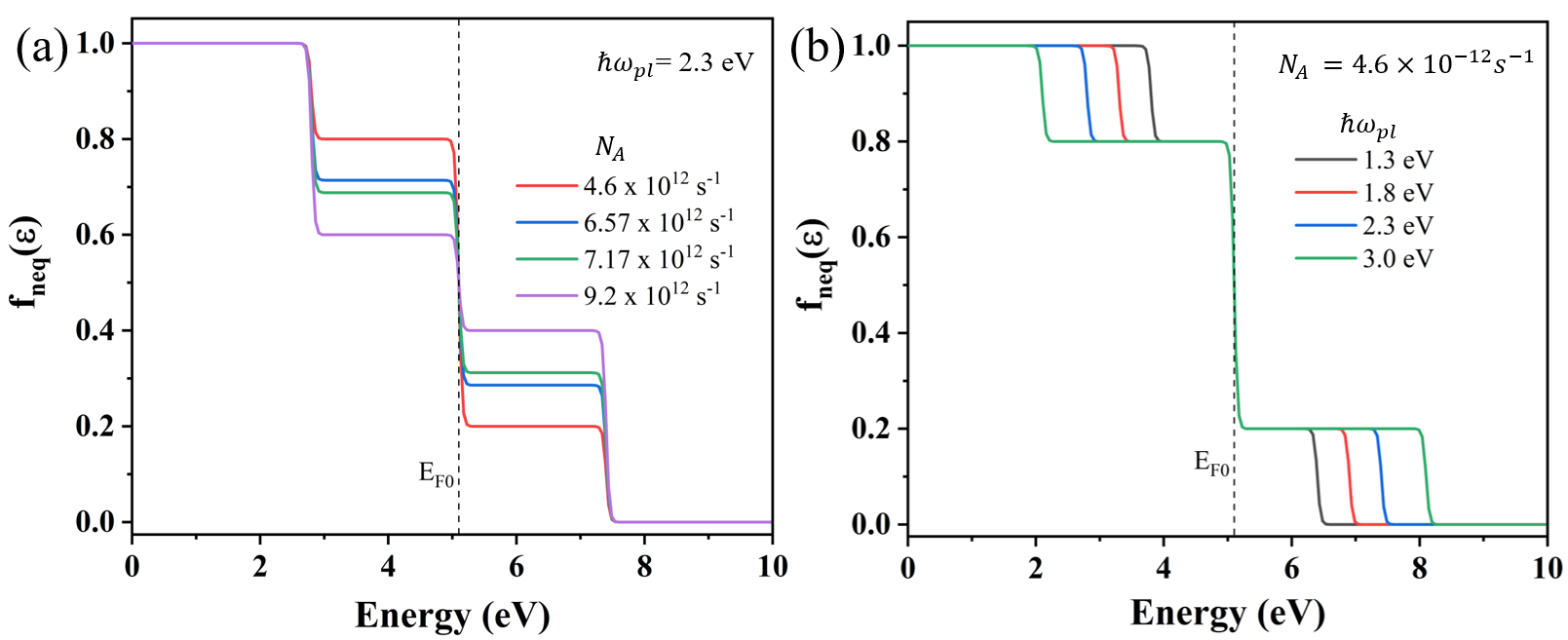}
	\caption{Hot electron distribution plotted for (a) various values of $N_A$ at fixed  $\hbar \omega_{pl}$ and (b) different excitation energies for constant $N_A$.   }
	\label{Fig:distribution}
\end{figure}

Localized surface plasmons excited in nanoparticles with large free electron densities typically dephases within ~10 fs. For smaller particles the LSPs decay via electron - hole pair generation. For gold nanoparticles with the LSP resonance in the visible regime, the electrons in the \textit{sp} band can transition from below the $E_F$ to empty states above in an intraband process, satisfying momemtum  conservation. The difference momentum may be provided by either phonons or electron  scattering \cite{Khurgin2019} generating low energy electron-hole pairs  \cite{Zheng2015}. Higher energy electron-hole pairs with energies equivalent to incident photons are generated when a spectrum of momentum values is provided by the plasmonic structure factor or by surface collisions. The hot carriers are short lived and thermalize via scattering with other hot carriers and phonons, a process which takes upto 100s of ps. However in case of CW excitation, the population of hot carriers is continuously replenished and reach a steady state distribution. These uncorrelated hot electron - hole pairs distribution is highly nonequilibrium and a large body of research has contributed to understanding the plasmon-induced hot carrier generation both from theoretical as well as experimental perspective \cite{Mertens2017a,Brown2016,Besteiro2017a,Besteiro2016,Bernardi2015b}.  
Here, we employ a modified semi-classical calculation \cite{Kornbluth2013a} to extract analytical expressions for hot electron distribution for a single nanoparticle under CW excitation in the steady state. We start by writing a joint probability density of plasmon absorption of frequency between $\omega$ and $\omega$ + $d\omega$  and excitation of an electron-hole pair at energies $\left(\epsilon, \epsilon -\hbar \omega\right)$ between $\epsilon$ and $\epsilon$ + $d\epsilon$,
\begin{equation}
	A\left(\epsilon_{}, \omega\right)= \frac{ n_{A}(\omega) D_{}( \epsilon,  \epsilon -\hbar \omega; k, k - \Delta k) g( \epsilon,  \epsilon-\hbar \omega)}{\int n_{A}(\omega)  D_{}(  \epsilon, \epsilon-\hbar \omega) g(\epsilon, \epsilon-\hbar \omega) d \epsilon d\omega}
\end{equation}
Here, $D\left(\epsilon, \epsilon-\hbar \omega ; k, k - \Delta k \right)$ is the squared magnitude of transition matrix element for $\epsilon_{j} \rightarrow \epsilon_{k} = \epsilon_{j }+\hbar \omega$ and $k_{j} \rightarrow k_{k} = k_{j} + \Delta k$ and $D$ is hereafter assumed to be 1. $g\left(\epsilon, \epsilon-\hbar \omega\right)$ is the product of free electron density of states $\rho(\epsilon)$ with the equilibrium Fermi distribution $f_{eq} (\epsilon)$. The general form of $g\left(\epsilon, \epsilon-\hbar \omega\right)$  is given by,
\begin{equation}
	g_{}\left(\epsilon_{k}, \epsilon_{j}\right)=\left\{f_{eq} \left(\epsilon_{j}\right) \rho\left(\epsilon_{j}\right)\right\} \times\left\{\left[1-f_{eq} \left(\epsilon_{k}\right)\right] \rho\left(\epsilon_{k}\right)\right\}
	\label{eq2}
\end{equation}
$n_{A}(\omega)$ is the number plasmons absorbed between $\omega$ and $\omega$ + $d\omega$ per unit time, per unit volume. $ N_{A}=\int d \omega n_{A}(\omega)$ is then \textbf{total number density of absorbed plasmons/time}.
Next we determine $N_{A}$ for a monochromatic, CW illumination with energy $\hbar \omega_{pl}$ (=$\hbar \omega_{vis}$).
\begin{eqnarray*}
	N_{A} &=& \frac{E_{plasmon}}{\hbar \omega_{pl} \times time}\\
N_{A}&=&\frac{ \left \langle P_{SP}\right \rangle}{\hbar \omega_{pl}}
\end{eqnarray*}
where, $E_{plasmon}$ is the total energy supplied by the absorbed plasmons. Neglecting other channels of energy loss, the average power dissipated $ \left \langle P_{SP}\right \rangle$ in the system that may be expressed in terms of the energy density of the plasmon field as,
\begin{equation}
	\left \langle P_{SP}\right \rangle= \omega_{pl} \textbf{Im}(\epsilon_{m}) \left \langle|E(x,t)|^{2}\right \rangle_{}
\end{equation}
here, $Im(\epsilon_{m})$ is the imaginary part of the nanoparticle's dielectric function $\epsilon_{m}$ and $|E(t)|^2$ is the squared net electric field. $N_A$ for a plasmonic nanosphere embedded in a dielectric material of permittivity $\epsilon_0$, excited at its resonant frequency is given by,
\begin{equation}
	N_A = \frac{\textbf{Im}(\epsilon_{m})\omega_{pl}}{2} \big(\frac{3\epsilon_0}{2\epsilon_{0}+\epsilon_{m}}\big)^2 |E|^2	
\end{equation}
where $|E|^2 $ is the excitation field intensity. $N_A$ contains all information about the metal nanoparticle,  it's environment and the plasmonically enhanced near field that ultimately determines the efficiency of hot carrier generation.  
Thus, the net change in electron population at any energy $\epsilon$ per unit time is given by 
$\frac{N_{A} }{\rho(\epsilon)}\phi(\epsilon)$, where $\phi(\epsilon)$  denotes the residual fraction of electrons at a given energy, and is calculated as; 
\begin{equation}
	\phi(\epsilon)= \int d\omega [ A\left(\omega_{}, \epsilon\right)- A\left(\omega_{}, \epsilon +\hbar\omega_{pl}\right)]
	\label{eq6}
\end{equation}
The hot electron excitation term $\frac{N_{A} }{\rho(\epsilon)}\phi(\epsilon)$ is compensated by electron thermalization occurring on the time scale $\tau$. Invoking relaxation time approximation, rate of change of electron population is written as,  
\begin{equation}
	\frac{d \Delta f_{neq}}{dt} = - \frac{\Delta f_{neq}}{\tau} + \frac{N_{A} }{\rho(\epsilon)}\phi(\epsilon) 
\end{equation}
where $\tau$ is the relaxation time constant. At steady state,  $\frac{d \Delta f_{neq}}{dt} = 0 $, so that $\Delta  f_{neq}$ becomes,  
\begin{equation}
	\Delta f_{neq}(\epsilon) = \frac{N_{A} \tau}{\rho(\epsilon)}\phi(\epsilon)
	\label{eq8}
\end{equation} Under CW illumination, the hot carrier population remains driven out of equilibrium and is continuously replenished, even in the presence of relaxation pathways such as electron-electron or electron-phonon scattering. Nevertheless, to estimate an upper bound for the hot electron generation rate, we can take the relaxation time to be 10 fs \cite{Kornbluth2013a,RN722,vadai2016plasmon}. Additionally, $\rho(\epsilon)$ can be taken to be energy independent at $0.1 eV^{-1}$ per atom for energies at a few eV from the Fermi level \cite{vadai2016plasmon,RN722}.
Combining equations \ref{eq2}, \ref{eq6} and \ref{eq8} we obtain the expression for nonequilibrium electron distribution as; 
\begin{equation}
		f_{neq}(\epsilon)= f_{eq} + \frac{N_{A} \tau}{\rho \hspace{0.03cm}\hbar \omega_{pl}} \tanh \left(\frac{(\epsilon-E_{F0}) }{2}\beta\right) \times     \frac{1}{1+2e^{-\hbar \omega_{pl} \beta}\cosh \left((\epsilon-E_{F0}) \beta\right)}
		\label{fneq}
\end{equation}
In calculating $f_{neq}(\epsilon), $ $\hbar \omega_{pl}$ = 2.3 eV which corresponds to the maximum emission intensity  in the ZnO PL spectrum. Figure \ref{Fig:distribution}(a) and (b) plots the variation in hot electron distribution $f_{neq}(\epsilon)$ as a function of $N_A$ as well as the excitation energy $\hbar \omega_{pl}$ at T = 300 K and $E_{F0} = 5.1 eV$ . The selected values of $N_A$ are motivated by the experimental NBE intensity increment factor of AZn samples, where maximum $N_A$ value of $9.2\times10^{12} s^{-1}$ corresponds to the 5AZn sample.
\newpage
\section*{S6. Photoluminescence lifetimes}
\begin{table*}[!h]
	\centering
	\begin{tabular}{ | m{1.5cm} | m{3.2cm} | m{5.2cm} | m{5.2cm} | m{1.5cm} |} 
		
		\hline
		\textbf{Material}  & \textbf{Plasmonic Nanostructure Geometry (SPR Energy)} & \textbf{Near band edge emission lifetimes} & \textbf{Defect level emission lifetimes}  &\textbf{Reference}\\ 
		\hline
		\hline
		ZnO thin film&\makecell{ Ag-NP array  \\ (3.1 eV) }& \makecell{$\tau_{0}$ - 1.49 ns (ZnO) \\ $\tau_{0}$ - 1.24 ns (ZnO-SP) }&\centering-&\cite{Wang2021}\\ 
		\hline
		ZnO NP&\centering Au-NP(2.5 eV)&\makecell{  $\tau_{NR}$ - 85 ps, $\tau_{R}$ - 21.9 ns (ZnO) \\ \\ $\tau_{NR}$ - 69 ps, $\tau_{R}$ - 20 ns (ZnO-SP)} & \centering- & \cite{RN680}\\
		\hline
		ZnO nanorod&\centering Al-NP (3 eV)&\makecell{$\tau_{0}$ - 161 ps (ZnO) \\$\tau_{0}$ - 53 ps (ZnO-SP)} & \centering- & \cite{RN681}\\
		\hline
		Indium-ZnO thin film    & Ag thin film (3.11 eV) &\centering-& \makecell{ $\tau_{0}$ - 1.6 ns (ZnO),\\$\tau_{R}$ - 2.5 ns, $\tau_{NR}$ - 110 ps    (ZnO-SP)}  & \cite{Kiba2016}\\
		\hline
		ZnO nanorod &\centering Au-NP(2.25 eV) &\makecell{ $\tau_{NR}$ - 25 ps, $\tau_{R}$ - 129 ps (ZnO) \\ \\$\tau_{NR}$ - 21 ps, $\tau_{R}$ - 91 ps    (ZnO-SP)}&\centering-  & \cite{Fiedler:20}\\
		\hline
		ZnO NP   & \centering Au-NP(2.25 eV) &\makecell{ $\tau_{NR}$ - 12 ps, $\tau_{R}$ - 1.75 ns (ZnO) \\ \\$\tau_{NR}$ - 19 ps, $\tau_{R}$ - 1.05 ns    (ZnO-SP)}&\centering-  & \cite{guidelli2015enhanced}\\
		
		\hline
	\end{tabular}
	\caption{ List of spontaneous emission lifetimes for the NBE and defect level emissions for various ZnO - plasmonic metal hybrids. Time constant generated from single exponential fit of PL decay curve is labeled as $\tau_{0}$. Bi-exponential fit of PL decay curve generates two time constants, labeled as $\tau_{NR}$ and  $\tau_{R}$.}
	\label{tab:lifetimes}
\end{table*}

\clearpage

\bibliography{SMAuZnOref}